\documentclass[twoside,12pt]{article}
\usepackage{epsfig}
\def\Journal#1#2#3#4{{#1} {#2} (#4) #3 }

\def\NPA{{\em Nucl. Phys.} A}

\def\NPB{{\em Nucl. Phys.} B}

\def\PLB{{\em Phys. Lett.} B}

\def\PRL{\em Phys. Rev. Lett.}

\def\PREP{\em Phys. Rep.}

\def\PRD{{\em Phys. Rev.} D}

\def\ZPC{{\em Z. Phys.} C}

\def\EPJC{{\em Eur. Phys. J.} C}

\def\R{\vec R}
\def\p{\vec p}

\def\q{\vec q}

\newcommand{\be}{\begin{equation}}
\newcommand{\ee}{\end{equation}}
\newcommand{\bea}{\begin{eqnarray}}
\newcommand{\eea}{\end{eqnarray}}

\def\z0{\rm Z^0}
\newcommand{\as}{\alpha_{\rm s}}

\newcommand{\oaa}{{\cal O}(\as^2)}
\newcommand{\oaaa}{{\cal O}(\as^3)}
\newcommand{\epem}{\rm e^+\rm e^-}
\newcommand{\yc}{y_{\rm cut}}
\newcommand{\amz}{\as(M_{\rm Z^0})}

\def\mz{M_{\rm Z^0}}

\def\d2{D_2}
\def\oq{\char'134}
\def\lamsb{\Lambda_{\overline{MS}}}
\def\ecm{E_{cm}}

\def\m2{\mu^2}
\def\q{\rm q}

\def\p{\rm p}

\def\q2{Q^2}
\def\asmu{\as (\mu^2)}
\def\asq{\as (\q2 )}

\def\R{\cal{R}}
\def\msbar{\overline{\mbox{MS}}}

\begin{document}

\title{ \vspace{1cm} Experimental Tests of Asymptotic Freedom}
\author{S. Bethke \\ Max-Planck-Institut f\"ur Physik, Munich, Germany}
\maketitle
\begin{abstract} 
Quantum Chromodynamics (QCD), the gauge field theory of the Strong 
Interaction, has specific features, asymptotic freedom and confinement,
which determine the behaviour of
quarks and gluons in particle reactions
at high and at low energy scales. 
QCD predicts that the strong
coupling strength $\as$ decreases with increasing energy 
or momentum transfer, and 
vanishes at asymptotically high energies.
In this review, the history and 
the status of experimental tests of
asymptotic freedom are summarised.
The world summary of measurements of $\as$ is updated, leading
to an unambiguous verification of the running of $\as$ and of 
asymptotic freedom, in excellent agreement with the predictions
of QCD.
Averaging a set
of measurements balanced between different particle processes
and the available energy range, results in a new
and improved world average of
$\amz = 0.1189 \pm 0.0010\ .$
\end{abstract}
%
\begin{flushright}
MPP-2006-54 \\
June 2006
\end{flushright}
%

\eject
\tableofcontents
\eject

\section{Introduction}
The Strong Interaction, binding quarks and gluons inside hadrons, 
is the strongest of the four fundamental forces in nature which we know today.
It is about 100 times 
stronger than the electromagnetic force, a factor of $10^{14}$ stronger than the
Weak Interaction, and a stunning factor of $10^{40}$ 
stronger\footnote{
40 orders of magnitude correspond to the difference in size of our entire universe 
to the size of an atomic nucleus.}
than the Gravitational Force, calculated
for two quarks at a distance of order 1~fm.
 
In spite of these differences, it is the Gravitational and the
Electromagnetic Forces which seem to play the dominating role in
the universe which we experience in the macroscopic world.
The reason for this wondrous fact is that the Strong Force, 
as well as the Weak Interaction, only acts at subatomic distances.
Taking the existence of atomic nuclei, their composition and
their masses for granted,
restricting ourselves to energies and temperatures below 
1~MeV and, respectively, $10^8$ Kelvin, the Strong Interaction is 
disguised in the description of the world.

At subatomic distances, however, this picture changes completely.
The Strong Force not only determines the binding of quarks and 
gluons inside hadrons, it also determines the cohesion
of protons and neutrons inside atomic nuclei.
Hadrons like protons and neutrons are responsible for more than 99\% 
of the mass of all visible matter in our universe, and those masses are 
mainly generated by the strong binding of quarks inside hadrons,
rather than by the (generally small) masses of the quarks themselves.

The restriction of the Strong Force to subatomic distances is a consequence
of two characteristic features: they are
called \oq Confinement" and \oq Asymptotic Freedom".

Confinement is a necessary requirement to explain the fact
that no isolated quarks have ever been observed in any experiment, although
symmetry arguments and scattering experiments in the 1960's established
quarks, with -1/3 or +2/3 of electrical charge units and a newly introduced
quantum property called "colour charge", as the basic constituents of hadrons.
Confinement determines that at large distances, or - equivalently - at low
momentum or energy transfer in elementary particle reactions, the 
Strong Force prevents the existence of free quarks:
Trying to separate two quarks from each other, for instance in
high energy scattering reactions, apparently results in an increase
of the force field's energy at large distances,
such that new quarks are created out of 
the vacuum -
the initial quarks \oq dress" up with other quarks to build hadrons.
These hadrons exhibit no net colour charge to the outside, such that 
they appear as elementary entities rather than the quarks 
themselves\footnote{
Van-der-Vaals-like remnant forces  bind protons and 
neutrons in atomic nuclei.}.

The term \oq Asymptotic Freedom" is used to describe the behaviour 
of quarks at
high energy or momentum transfers, or - equivalently - at small distances.
Also this feature is based on experimental observations:
In high energy scattering processes between leptons (e.g. electrons or neutrinos) 
with protons or neutrons, the dynamics reveal that scattering occurs at 
pointlike and massless constituents, the quarks, rather than at a homogenuous 
object with the size of a proton. 
Apparently, at sufficiently high momentum transfers, quarks behave
like free or weakly bound
particles.
Also, quarks knocked out of a hadron, in a high energy scattering process, 
were never observed as free particles.
Instead, they emerge as dressed-up hadrons
or bundles (jets) of hadrons escaping from the interaction region.

The fact that the strong interaction becomes \oq weak" at 
high energy scales, and vanishes to zero at asymptotically high energies,
led to the term  \oq Asymptotic Freedom". 
Any theory describing and predicting the dynamics of quarks inside hadrons
and in high energy reactions therefore had to satisfy and 
include these two
extremes: Confinement, also called \oq infrared slavery",
and Asymptotic Freedom.

A theoretical description of the Strong Interaction 
by a consistent quantum field
theory, called Quantum Chromodynamics (QCD),
was presented in 1972 by Fritzsch and Gell-Mann \cite{fritzsch},
which was formaly published later in \cite{fgl}.
At the same time it was found by Gross, Politzer and Wilczek
that non-Abelian gauge theories, such as
QCD, exhibit the propoerty of Asymptotic Freedom \cite{qcd}.
This discovery was honored with the Nobel Price in 2004, 
30 years after the original findings.

Theoretical breakthroughs and discoveries, before being honored by the 
Nobel Price, must be established by experimental measurements.
The price for Asymptotic Freedom therefore also had to await 
experimental verification.
In this article, the long but exciting road to experimental evidence for 
Asymptotic Freedom shall be summarised.

\section{Historical Development}

\subsection{A New Force}

In the 1930's,
protons and Neutrons were recognised as the building blocks of atomic nuclei. 
A new and elementary force, the so-called Strong Force, was introduced 
phenomenologically, in order to explain the binding of these particles 
inside nuclei.
These binding energies were known to be much larger than what could be 
expected from the well-known electromagnetic force.
Also, the electromagnetic force was known to have infinite range while 
the Strong Force apparently acted only within nuclear distances of the
order of femtometer.

Shortly after these perceptions, Hideki Yukawa argued that the 
range of the electromagnetic force was infinite because the associated exchange
particle, the photon, was massless, and proposed that the short range of the 
strong force was due to the exchange of a massive particle which was called
\oq meson" \cite{yukawa}.
From quantum mechanical arguments and from Heisenberg's uncertainty principle
Yukawa predicted a meson mass\footnote{
From now on, in this review, a
system of units is utilized where the speed of light and Planck's constant are set
to unity, $c = \hbar = 1$, such that energies, momenta and masses are all given in
units of eV.}
of about 100~MeV/c$^2$.
At that time, no particle with a mass in that range was known.

Very soon after, in 1937, a new particle was dicovered whose mass was very close
to that of Yukawa's prediction of the meson - however, this particle 
in fact was
the muon which turned out not to be
subject to the strong interaction.
Finally,
in 1947 strongly interacting particles with masses close to
Yukawa's prediction were found \cite{powell} in emulsion experiments at 
high altitudes.
A particle which met Yukawa's requirements was found, and it
was called \oq pion".
Although today we know that pions, like protons and neutrons, are composite
particles made from quarks, and the Strong Interaction is not mediated
through the exchange of pions, 
Yukawa's theory stimulated major advances in the understanding and description
of the strong interaction.

\subsection{Quarks}

In the decades after World War II, the number of strongly interacting particles
exploded - which was basically due to the development of particle
accelerators with steadily increasing energies.
Strongly interacting particles were then called \oq hadrons",
which differentiate into mesons (with integer or zero spins) and baryons 
(with spins of 1/2 or 3/2).
Gell-Mann and Zweig realised in 1964 that the whole spectroscopy of
hadrons could be explained by a small number of \oq quarks"
if baryons are made out of 3 quarks, 
and mesons out of one quark plus one antiquark
\cite{zweig}.
Those quarks, initially 3 different species, then must have 1/3 or 2/3
of the elementary charge unit, and they must have spin 1/2.

\begin{figure}[ht]
\begin{center}
\epsfxsize14.0cm\epsffile{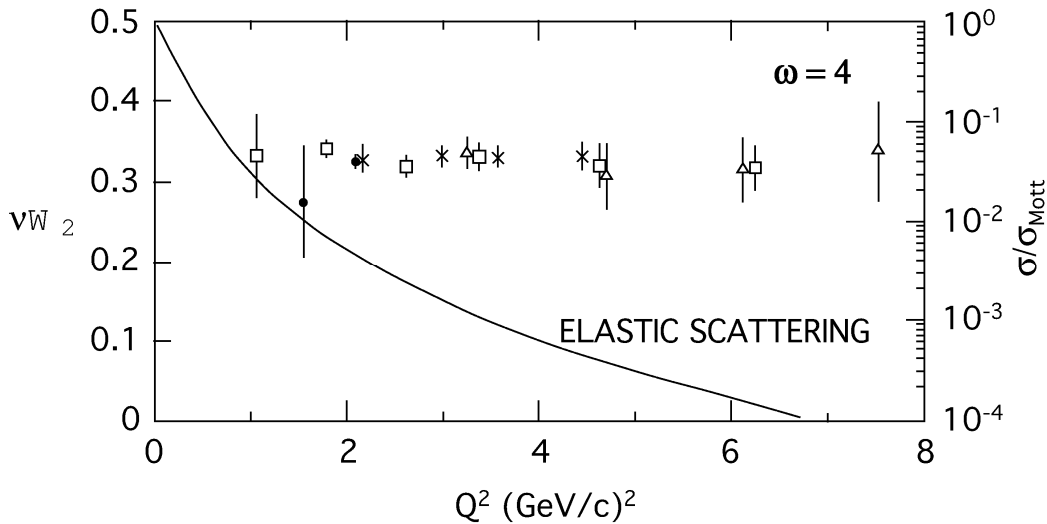} 
\end{center}
\caption{The structure function of the proton, describing the electron-
proton scattering in units of the Mott cross section, i.e. the generalised
Rutherford cross section at high energies, as
measured at SLAC~\cite{slac-data}.
$\omega = 4$ corresponds to $x = 0.25$ in figure~\ref{fig:f2}.
\label{fig:slac}}
\end{figure}

By the end of the 1960's, the static picture of quarks as the
constituents of hadrons was confirmed through the dynamics observed
at high energy electron-proton scattering experiments at the
Stanford Linear Accelerator (SLAC)\cite{slac-data},
summarised in figure~\ref{fig:slac}:
the cross sections - instead of decreasing with increasing momentum
transfer as expected for elastic scattering of electrons at protons
as a whole - showed a \oq scaling" behaviour, as it should occur if the 
electrons scatter on quasi-free, pointlike 
and nearly massless constituents inside the protons.

While the quark model was very successful in describing the properties,
multitude and dynamic behaviour of hadrons, 
it had severe shortcomings
such as an obvious violation of the Pauli-principle for hadrons containing
two or three identical quarks in 
relative S-waves and in identical spin states, the prediction of
the neutral pion's lifetime which was wrong by a factor of nine, and the fact that
no particles (quarks) with 1/3 or 2/3 of the elementary charge unit could ever be 
observed at particle colliders.

\subsection{Chromo-Dynamics}

To overcome these shortcomings, the idea that the strong force 
couples to a new quantum poperty, called colour charge 
\cite{greenberg,nambu}, in analogy to the photon
in QED which couples to the electric charges. 

The introduction of the new quantum number, describing 3 
different colour quantum states of each of the quark species,
solved the puzzle of spin-statistics, saved the Pauli-principle and
explained the missing factor of nine ($=3^2$) of the pion lifetime.
The notion that hadrons consist either of three quarks (baryons like the proton)
or a quark and an antiquark (mesons), 
arranged such that the net color charge
of the hadron would vanish - i.e. \oq white" as a superposition of the three 
elementary color states or of a particular color and its anti-color - could
account for the fact that the strong force is short-ranged.

Finally, in the early 1970's, a field theory of the strong force,
Quantum Chromodynamics (QCD), was
introduced and presented in canonical form \cite{fritzsch,fgl}.
Here, 
coloured spin-1 particles called \oq gluons", which - in contrast to the 
case of photons in QED - carry colour charge themselves, couple to the
colour charges of quarks, and also to coloured gluons themselves.
Chromo-Statics turned into Chromo-Dynamics.

\subsection{Asymptotic Freedom}

A few years before, it was
demonstrated in model theories that charges may change their effective size
when they are probed in scattering experiments, at large and at small distances.
These changes were described through Symanzik's $\beta$-function 
\cite{symanzik}, see equation~2 below:
the effective size of the coupling strength is a function of the energy or
momentum transfer, $\q2$.
The SLAC data on approximate scaling of the proton structure function,
c.f. figure~\ref{fig:slac}, and the notion of free quarks
inside the proton required a negative $\beta$-function.
At that time, however, all field theories probed so far
had a positive $\beta$-function.

Gross, Politzer and Wilczek finally demonstrated in 1973 that Chromo-Dynamics, 
with coloured quarks and gluons, obeying SU(3)$_{colour}$ symmetry,
generated a negative $\beta$-function \cite{qcd}, 
i.e. that quarks and gluons are asymptotically free.
This explained the SLAC data, and at the same time leads - for small momentum
transfers or at large distances, to an increase of the coupling strength, thus 
motivating confinement.

Another important consequence from asymptotic freedom is the fact that
the strong coupling $\as$ is small enough, at sufficiently large 
momentum transfers, to allow application of perturbation theory in order
to provide quantitative predictions of physical processes.

With this discovery Quantum Chromodynamics started its triumphal procession 
as being $the$ field theory of the strong interaction.
It was, however, a long and difficult path before QCD was commonly
accepted to be exactly that.
Many refined calculations, theoretical predictions and experimental verifications were ventured.
The most important features of QCD, asymptotic freedom and/or,
equivalently, the existence of colour charged gluons,
had to be tested, quantified and \oq proven" by experiment.
The size of the strong coupling parameter, $\asq$, had to be determined and 
its energy dependence verified to be compatible with asymptotic freedom.

\subsection{Scaling Violations}

The violation of approximate scaling of the proton structure function
was one of the first experimental signatures proposed to test QCD \cite{gross2}.
Deep inelastic electron-proton (or more generally, lepton-nucleon) scattering
processes basically depend on two kinematical parameters.
The usual choice of these parameters are
the momentum transfer $\q2$ between the lepton and the struck quark,
and the fraction $x$ of the proton's momentum which is carried by the
struck quark.

The structure function $F_2$ of the proton 
basically parametrises the population of quarks
with momentum fraction $x$, as a function of $\q2$: $F_2(x,\q2)$.
QCD predicts that the region of large $x$
is depopulated for higher momentum transfers, while the population at low
$x$ should increase with increasing $\q2$.

A simple argument for this behaviour can be given in terms of 
the equivalence of momentum transfer and physical resolution:
with higher $\q2$, smaller distances are probed. 
At smaller distances,
an increased radiation of soft (low energy) gluons should be resolved.
These gluons increase the population
of partons at small $x$, and - at the same time -
diminish the momentum fraction and the relative contribution
of quarks at large $x$.

On top of it's kinematical origin, scaling violations in $x$ and $Q^2$ are
modified by the specific energy dependence of the strong coupling.
Note that a qualitative observation of scaling violations of structure
functions alone does not yet prove Asymptotic Freedom - such a conclusion 
requires a large lever arm in $\q2$ and a precise
study of the functional form of observed scaling violations,
as e.g. the logarithmic slopes, $\partial F_2(x,Q^2) / \partial ln Q^2$. 

While scaling violations were already observed
in the 1970's,
initially it was not possible to conclude that asymptotic freedom was 
actually seen \cite{buras,duke}.
Instead, unknown higher order QCD terms, the effects of target masses 
and nonperturbative effects (\oq higher twists") could 
cause similar effects as those predicted by asymptotic freedom.
Only with the availability of much higher $\q2$ and the ability
to probe much smaller values of $x$, e.g. at the electron-proton collider HERA,
stricter conclusions were possible in the past 10 to 15 years.
These will be presented in more detail in section~5.

\subsection{Quark and gluon jets in high energy collisions}

Another class of predictions, basically governed by the concept of
confinement, was the
expectation of collimated hadron jets in high energy reactions
like $\epem \rightarrow q\bar{q}$ \cite{brandt,drell-jets-1970}.
Indeed, by 1975, the emergence of 2-jet structure 
was observed when increasing the $\epem$ center 
of mass energy $\ecm$ from 3 to 7.4~GeV \cite{firstjets},
confirming the basic ideas of the quark-parton model.

\begin{figure}[ht]
\begin{center}
\epsfxsize12.0cm\epsffile{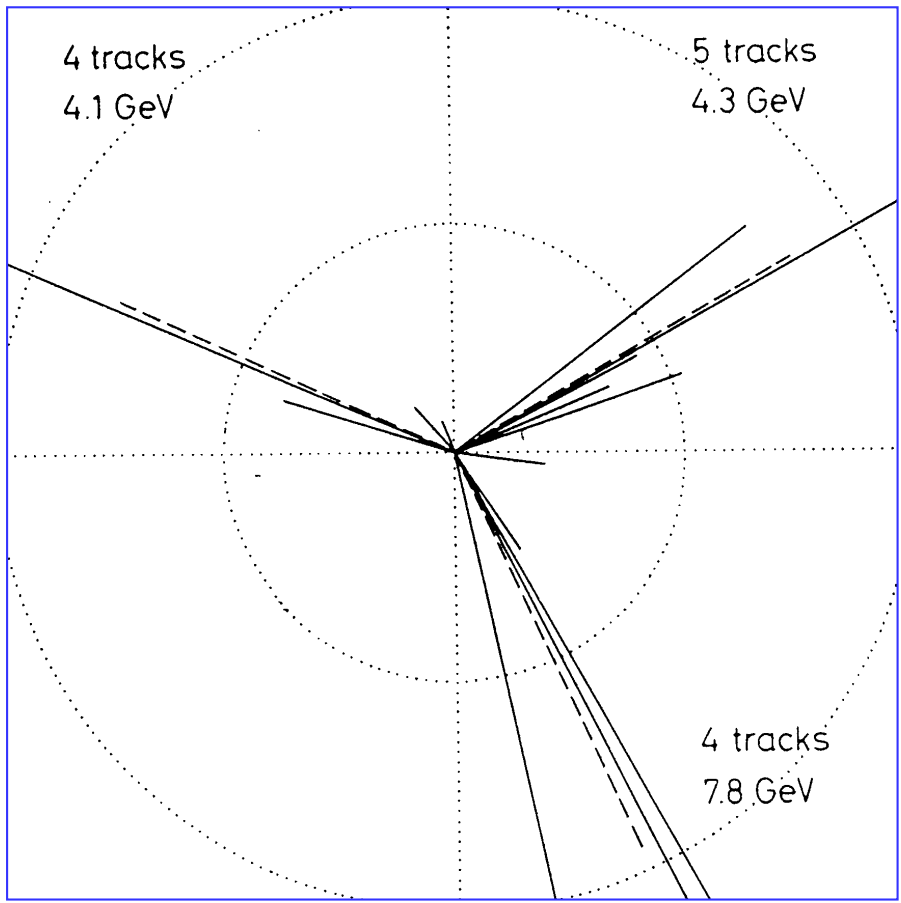} 
\end{center}
\caption{The first 3-jet event, observed by  the TASSO experiment
at the PETRA $\epem$ storage ring \cite{gluon}.
\label{fig:3-jet}}
\end{figure}

The radiation of energetic gluons off high energetic quarks was
predicted from QCD, and hence the emergence of a third hadron jet 
was expected in $\epem$ annihilation at higher energies \cite{gaillard}.
In 1979, 3-jet structures were observed at the PETRA $\epem$ collider at
$\ecm \approx 30$~GeV \cite{gluon}, see figure~\ref{fig:3-jet}.
The $3^{rd}$ jet could be attributed to the emission of a third parton
with zero electric charge and spin~1 \cite{gspin} -- the gluon
was discovered and explicitly seen!

After an almost parameter-free description of hadronic
event shapes \cite{hoyer,ali}
had served to establish gluon radiation beyond any
doubts, 
first determinations of the coupling strength $\as$
were performed.
More detailed studies of the jet structure and the 
shape of hadronic events provided insights into the color structure of
the gluon, like for instance
the \oq string effect" \cite{string-exp} which was expected
in QCD due to the higher color charge of the gluon \cite{string-th}.

\subsection{Status in the late 1980's}

\begin{figure}[ht]
\begin{center}
\epsfxsize12.0cm\epsffile{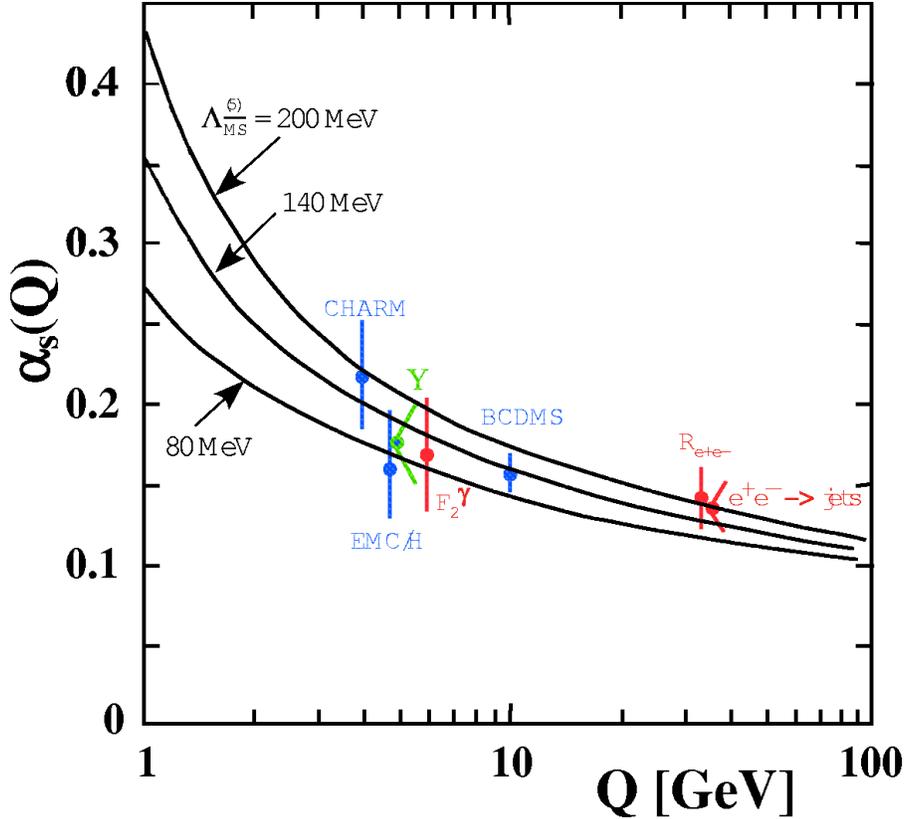} 
\end{center}
\caption{Summary of measurements of $\as$ in 1989 \cite{altarelli-89}.
Shown are results from various experiments in deep inelastic
lepton-nucleon scattering as well as combined results from $\epem$
collisions, together with the QCD expectation
of a running $\as$ for different values of $\lamsb$ (see 
section~3).
\label{fig:as89}}
\end{figure}

Towards the end of the 1980's, after successful exploitation of the 
experimental programs of
the $\epem$ storage rings PETRA at DESY and PEP at SLAC, 
the proton-antiproton collider at CERN
and many deep inelastic lepton-nucleon experiments at CERN, at
Fermilab and at SLAC, \oq {\em the experimental support for QCD is quite
solid and quantitative}" \cite{altarelli-89}, however convincing $proofs$ of the
key features of QCD, of the gluon-selfcoupling and/or of asymptotic freedom,
were still not available.
As an example, the summary of measurements of $\as$ at different energy scales
in 1989, as reproduced in figure~\ref{fig:as89} 
\cite{altarelli-89}, was compatible with the QCD expectation of the
running of $\as$, but did not yet allow to draw more concrete conclusions.

Foundations were layed, however, to prepare for more quantitative tests 
with the upcoming higher energy colliders, like the $\epem$ colliders 
LEP at CERN and SLC at SLAC, the Tevatron $p\bar{p}$ collider at Fermilab and
the HERA electron-proton collider at DESY, who all started operation 
in the time from 1987 to 1991:

\begin{itemize}
\item 
From the summary of $\as$ measurements shown in figure~\ref{fig:as89},
 using the QCD prediction of a running coupling,
 the value of $\as (Q^2 = \mz^2)$, at the energy scale of the $\z0$ boson mass,
$\mz \approx 90~$GeV, was predicted to be $\amz = 0.11 \pm 0.01$.
Experiments at LEP and SLC were determined to scrutinise this prediction with
high accuracy \cite{altarelli-89}.

\item
Deep inelastic scattering results, extending to much higher values
of $Q^2$ and much lower values of $x$, should
proove scaling violations of nucleon structure functions.

\item
Significant progress in jet physics allowed to prepare 
for direct tests of asymptotic freedom, through the energy dependence
of jet production rates \cite{jadejet1,jadejet2}, and of the gluon-selfcoupling,
through spin-correlations in 4-jet hadronic final states in $\epem$
annihilation \cite{sbzerwas}.

\item 
Precise determinations of $\as$ 
at low and at high energy scales, in $\epem$ annihilation,
at hadron colliders and in deep inelastic lepton-nucleon collisions were
expected to emerge, with the prospect of proving the energy dependence
of $\as$, and thus, asymptotic freedom.

\item
Theoretical developments like precise calculations in next-to-leading (NLO)
and next-next-to-leading order (NNLO) pertubation theory,
advanced methods of handling theoretical uncertainties, and improvement
in the understanding of the nonperturbative hadronisation process, in terms of advanced Monte Carlo
models, had started and were expected to be confronted
with future high precision experimental data.

\end{itemize}

After a short summary of the theoretical basics of QCD in section~3, the
developments of experimental test of QCD, and in particular of
asymptotic freedom,
following the status as summarised
above, will be presented in more detail in sections 4 to 6.

\section{Theoretical Basis}

The concepts of QCD are portrayed in many text books and articles, se e.g.
\cite{ellis-book,collins-book,muta-book,pdg,concise},
and in particular, the article of G.M.Prosperi, M. Raciti and C.Simolo
in this issue of PPNP \cite{PRS}.
In the following, a brief summary of the basics of perturbative QCD
and of the energy dependent strong coupling parameter, $\as$, is given.

\subsection{Renormalization}  

In quantum field theories like 
Quantum Chromodynamics (QCD) and Quantum Electrodynamics (QED),
physical quantities $\cal{R}$ 
can be expressed by a perturbation series in powers of the
coupling parameter $\as$ or $\alpha$, respectively.
If these couplings are sufficiently small, i.e. if
$\as \ll 1$, the series may converge sufficiently quickly such that it
provides a realistic prediction of $\cal{R}$ 
even if only a limited number of perturbative orders 
will be known. 

In QCD, examples of such quantities are cross sections, decay rates,
jet production rates or hadronic event shapes.
Consider $\cal{R}$ being dimensionless and
depending on $\as$ and on a single energy scale $Q$.
This scale
shall be larger than any other relevant, dimensional parameter such as quark
masses. 
In the following, these masses are therefore set to zero.

When calculating $\cal{R}$ as a perturbation series 
of a pointlike field theory in $\as$, ultraviolet
divergencies occur.
These divergencies are removed by the \oq renormalisation"
of a small set of physical parameters.
Fixing these parameters at a given scale and absorbing this way 
the ultraviolet divergencies, introduces a second but artificial
momentum or energy scale $\mu$.
As a consequence of this procedure, $\cal{R}$ and $\as$ become functions of
the renormalization scale $\mu$.
Since $\cal{R}$ is  dimensionless, we assume that it only depends on the ratio
$Q^2 / \mu^2 $ and on the renormalized coupling $\as (\mu^2 )$:
$$ {\cal R} \equiv {\cal R}(Q^2 / \mu^2, \as );\ \as \equiv \as (\mu^2). $$

Because the choice of $\mu$ is arbitrary, however, 
the actual value of the experimental observable $\cal{R}$ cannot depend
on $\mu$, so that

\begin{equation} \label{eq-muindependence}
 \mu^2 \frac{{\rm d}}{{\rm d} \mu^2} {\cal R} (Q^2 / \mu^2 , \as )
= \left( \mu^2 \frac{\partial }{\partial \mu^2 } + \mu^2 \frac{\partial
\as}{\partial \mu^2} \frac{\partial }{\partial \as } \right) {\cal R} 
=^{\hskip -5pt !}\ 0 \ ,
\end{equation}

\noindent where the derivative is multiplied with $\mu^2$
in order to keep the expression dimensionless. 
Equation~\ref{eq-muindependence} implies that any explicite dependence of
$\cal{R}$ on
$\mu$ must be cancelled by an appropriate $\mu$-dependence of $\as$
to all orders.
It would therefore be natural to identify the renormalization scale with the
physical energy scale of the process, $\mu^2 = Q^2$, eliminating the
uncomfortable presence of a second and unspecified scale. 
In this case, $\as$ transforms to the \oq running coupling constant"
$\asq$, and the energy dependence of $\cal{R}$ enters only 
through the energy dependence of $\asq$.

Any residual $\mu$-dependence is a measure of the quality of a given
calculation in finite perturbative order.

\subsection{$\as$ and its energy dependence}
While QCD does not predict the absolute size of $\as$, its energy dependence 
is precisely determined.
If $\as (\mu^2)$ is measured at a given
scale, QCD definitely predicts its size at
any other energy scale $Q^2$ through the renormalization group equation
\begin{equation} \label{eq-rge}
Q^2 \frac{\partial \asq}{\partial Q^2} = \beta \left( \asq \right) \ .
\end{equation}
\noindent The perturbative expansion of the $\beta$ function is calculated to
complete 4-loop approximation \cite{beta4loop}:
\begin{equation} \label{eq-betafunction}
\beta (\asq ) = - \beta_0 \as^2(Q^2) - \beta_1 \as^3(Q^2) - \beta_2 \as^4(Q^2) -
\beta_3
\as^5(Q^2)  + {\cal O}(\as^6)\ ,
\end{equation}
\noindent where
\begin{eqnarray} \label{eq-betas}
\beta_0 &=& \frac{33 - 2 N_f}{12 \pi}\ , \nonumber \\
\beta_1 &=& \frac{153 - 19 N_f}{24 \pi^2}\ , \nonumber \\
\beta_2 &=& \frac{77139 - 15099 N_f + 325 N_f^2}{3456 \pi^3}\ , \nonumber \\
\beta_3 &\approx & \frac{29243 - 6946.3 N_f + 405.089 N_f^2 + 1.49931 N_f^3}
        {256 \pi^4} \ ,
\end{eqnarray}
\noindent and $N_f$ is the number of active quark flavours at the energy
scale $Q$.
The numerical constants in equation~\ref{eq-betas} are functions of the group constants $C_A = N$ and $C_F = (N^2 -1) / 2N$, for theories exhibiting $SU(N)$
symmetry;   
for QCD and $SU(3)$,
$C_A = 3$ and $C_F = 4/3$.
$\beta_0$ and $\beta_1$ are independent of the
renormalization scheme, while all higher order $\beta$ coefficients are
scheme dependent.
\subsection{Asymptotic freedom and confinement}

A solution of equation~\ref{eq-rge} in 1-loop approximation, i.e.
neglecting
$\beta_1$ and higher order terms, is
\begin{equation} \label{eq-as1loop}
\asq = \frac{\as (\mu^2 )}{1 + \as (\mu^2 ) \beta_0 \ln{\frac{\q2}{\mu^2}}}\ .
\end{equation}
\noindent 
Apart from giving a relation between the values of $\as$ at
two different energy scales $Q^2$ and $\mu^2$,
equation~\ref{eq-as1loop} also demonstrates the property of asymptotic freedom:
if $\q2$ becomes large and $\beta_0$ is positive, i.e. if
$N_f < 17$, $\asq$ will asymptotically decrease to zero.

Likewise, equation~\ref{eq-as1loop} indicates that $\asq$ grows to large values 
and, in this perturbative form,
actually diverges to infinity at small $Q^2$: for instance, with
$\as ( \mu^2 \equiv M^2_{\rm Z^0}) = 0.12$ and for 
typical values of $N_f = 2\ ...\ 5$,
$\asq$ exceeds unity for $Q^2 \leq \cal{O} \rm{(100~MeV~...~1~GeV)}$.
Clearly, this is the region where perturbative expansions in $\as$ are not
meaningful anymore, and we may regard energy scales below
the order of 1~GeV as the nonperturbative region where confinement sets in, and
where equations~\ref{eq-rge} and~\ref{eq-as1loop} cannot be applied.

Including $\beta_1$ and higher order terms, similar but more complicated
relations for $\asq$, as a function of $\as (\mu^2 )$ and of
$\ln{\frac{\q2}{\mu^2}}$ as in equation~\ref{eq-as1loop}, emerge.
They can be solved numerically, such that for a given value of $\as (\mu^2 )$, 
choosing a suitable reference scale like the mass of the $\z0$ boson,
$\mu = M_{\z0}$,
$\asq$ can be accurately determined at any energy scale $\q2 \geq 1~\rm{GeV}^2$.

If we set
$$\Lambda^2 = \frac{\mu^2}{e^{1/\left( \beta_0 \as (\mu^2)\right) }}\ ,$$
\noindent
a dimensional parameter $\Lambda$ is introduced  such that 
equation~\ref{eq-as1loop} transforms into
\begin{equation} \label{eq-as1loop-2}
\asq = \frac{1}{\beta_0 \ln (\q2 / \Lambda^2)}\ .
\end{equation}
\noindent 
Hence, the $\Lambda$ parameter is
technically identical to the energy scale $Q$ where $\asq$ diverges to
infinity, $\asq \rightarrow \infty$ for $Q^2 \rightarrow \Lambda^2$.
To give a numerical example, $\Lambda \approx 0.1$~GeV for 
$\as (\mz \equiv 91.2\ \rm{GeV}) = 0.12$ and
$N_f$ = 5.

The parametrization of the running coupling $\asq$ with $\Lambda$ instead of 
$\as (\mu^2)$ has become a common standard, see e.g.~\cite{pdg}, and will
also be adopted here. 
While being a convenient and well-used choice, however, this parametrization has
several subtleties:

First, requiring that $\asq$ must be continuous when crossing
a quark threshold\footnote{Strictly speaking, {\em physical observables} $\R$ 
rather than $\as$ must be continuous, which may lead to small discontinuities
in $\asq$ at quark thresholds in finite order perturbation theory; see
section~3.7.},
$\Lambda$ actually depends on the number of active quark flavours. 
Secondly, $\Lambda$ depends on the renormalization scheme, see e.g. reference
\cite{collins-book}. 
In this review, the so-called \oq modified minimal
subtraction scheme" ($\msbar$)
\cite{msbar} will be adopted, which also has become a common standard~\cite{pdg}.
$\Lambda$ will therefore be labelled $\lamsb^{(N_f)}$ to indicate these
peculiarities.
\begin{figure}[ht]
\begin{center}
\epsfxsize12.0cm\epsffile{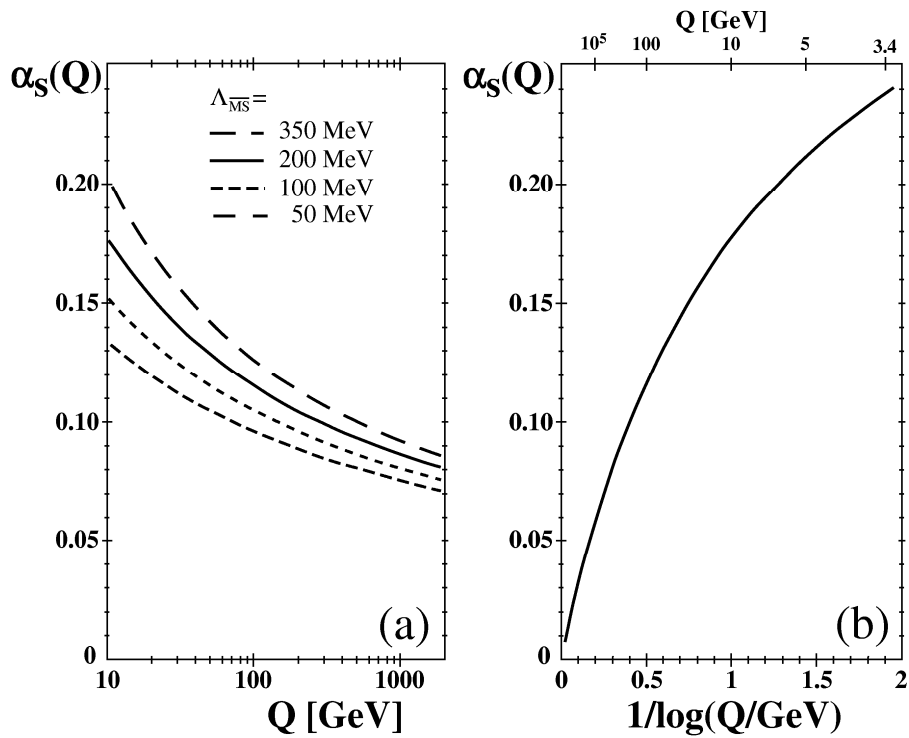} 
\end{center}
\caption{(a) The running of $\as (Q)$, according to
equation~\protect\ref{eq-as4loop},  in 4-loop approximation,
for different values of $\lamsb$;
(b) same as full line in (a), but as function of 1/log(Q/GeV)
to demonstrate asymptotic freedom, i.e. $\asq \rightarrow 0$ for
$Q \rightarrow \infty$.
\label{fig:asq-4L}}
\end{figure}

\subsection{The running coupling}

In complete 4-loop approximation and using the $\Lambda$-parametrization, the
running coupling is thus given \cite{alphas-4loop} by
\begin{eqnarray} \label{eq-as4loop}
\as (Q^2) &=& \frac{1}{\beta_0  L} 
               - \frac{1}{\beta_0^3 L^2} \beta_1 \ln  L  \nonumber \\
          &+& \frac{1}{\beta_0^3 L^3} \left( \frac{\beta_1^2}{\beta_0^2}
              \left( \ln^2 L - \ln L - 1 \right) + \frac{\beta_2}{\beta_0}
               \right) \nonumber \\
          &+& \frac{1}{\beta_0^4 L^4} \left( \frac{\beta_1^3}{\beta_0^3}
              \left( - \ln^3 L + \frac{5}{2} \ln^2 L + 2 \ln L - \frac{1}{2}
              \right) - 3 \frac{\beta_1 \beta_2}{\beta_0^2} \ln L
              + \frac{\beta_3}{2 \beta_0} \right)  \nonumber \\
\  \ 
\end{eqnarray}
\noindent
where $L = Q^ 2 / \lamsb^2 $.
The first line of equation~\ref{eq-as4loop} includes the 1- and the 2-loop
coefficients, the second line is the 3-loop and the third line is the 4-loop
correction, respectively.

\begin{figure}[ht]
\begin{center}
\epsfxsize12.0cm\epsffile{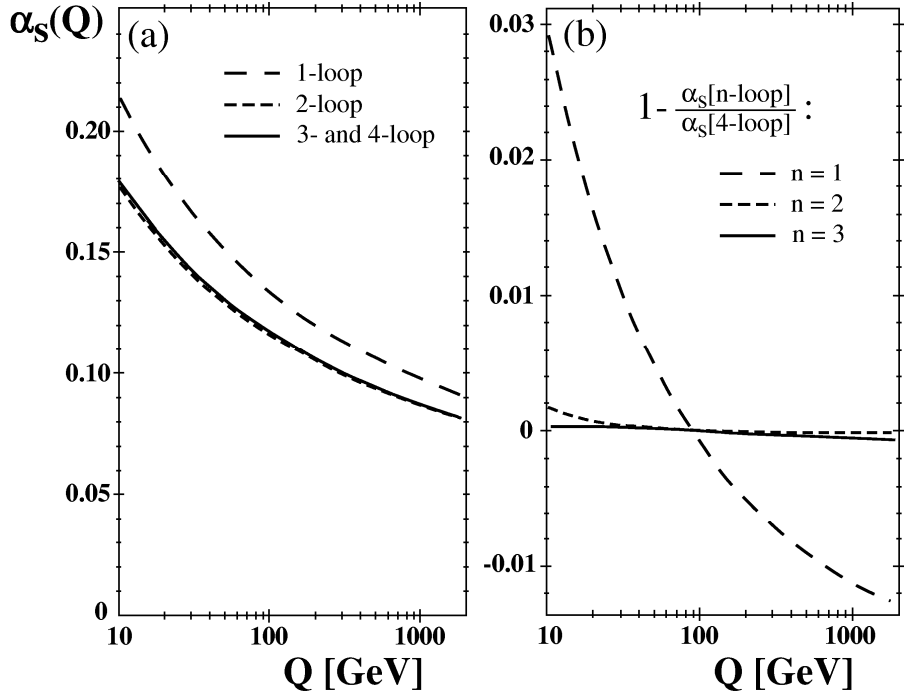} 
\end{center}
\caption{(a) The running of $\as (Q)$, according to
equation~\protect\ref{eq-as4loop},  in 1-,
2- and 3-loop approximation, for $N_f = 5$ and the same value of 
$\lamsb = 0.22$~GeV.
The 4-loop prediction is indistinguishable from the 3-loop curve. 
(b) Fractional difference between the 4-loop and the 1-, 2- and 3-loop
presentations of
$\as (Q)$, for $N_f = 5$ and $\lamsb$ chosen such that, in each order, $\amz
= 0.119$.
\label{fig:asq-orders}}
\end{figure}

The functional form of $\as (Q)$, for 4 different values of $\lamsb$ between 
50~MeV and 350 MeV, is diplayed in figure~\ref{fig:asq-4L}(a).
The slope and dependence on the actual value of $\lamsb$ 
is especially pronounced at small $\q2$, while at large $\q2$ both
the energy dependence and the dependence of $\lamsb$ becomes increasingly
feeble.
Nevertheless, $\as$ decreases with increasing energy scale and tends to zero
at asymptotically high $\q2$.
This is demonstrated in figure~\ref{fig:asq-4L}(b), where $\asq$ is
plotted as a function of $1/\log Q$.

Any experimental proof of asymptotic freedom will therefore require
precise determinations of $\as$, or of other observables which depend on $\asq$,
in a possibly large range of energy scales.
This range should include as small as possible energies, since the relative
energy dependence is largest there.
To date, precise experimental data and respective QCD analyses are available
in the range of $Q \approx 1$~GeV to a few hundred GeV.

\subsection{Relative size of finite order approximations}

The importance of higher order loop corrections and the degree of
convergence of the perturbative expansion for the
running $\as$, see equation~\ref{eq-as4loop}, is analysed 
and demonstrated in figure~\ref{fig:asq-orders}.
In part (a), the
1-, the 2- and the 3-loop approximation of equation~\ref{eq-as4loop}, each
with
$\Lambda = 0.220$~GeV, is shown.
As can be seen, there is an almost 15\% decrease of $\as$ when changing from 1-loop
to 2-loop approximation, for the same value of $\Lambda$.
The difference
between the 2-loop and the 3-loop prediction is only about 1-2\%, and it
is less than 0.01\% between the 3-loop and the 4-loop presentation which cannot
be resolved in the figure.

The fractional difference in the energy dependence of $\as$,
$\frac{(\as^{(4-loop)} - \as^{(n-loop)})}{\as^{(4-loop)}}$, for $n$~=~1,~2 and~3,
is presented in figure~\ref{fig:asq-orders}(b).
Here, in contrast to figure~\ref{fig:asq-orders}(a), the values of $\lamsb$ were
chosen such that $\amz = 0.119$ in each order,
i.e., $\lamsb = 93$~MeV (1-loop), $\lamsb = 239$~MeV (2-loop), and
$\lamsb = 220$~MeV (3- and 4-loop).
Only the 1-loop approximation shows sizeable differences of up to
several per cent, in the energy and parameter range chosen, while the 2- and
3-loop approximation already reproduce the energy dependence of the best, i.e.
4-loop, prediction quite accurately.

\subsection{Quark masses and thresholds}

So far in this discussion, finite quark masses $m_q$ were neglected, assuming that
both the physical and the renormalization scales $Q^2$ and $\mu^2$, respectively,
are larger than any other relevant energy or mass scale involved in the problem.
This is, however, not entirely correct, since there are several
QCD studies and $\as$ determinations at energy scales around
the charm- and bottom-quark masses of about 1.5 and 4.7~GeV, respectively.

Finite quark masses may have two major effects on actual QCD studies:
Firstly, quark masses will alter the perturbative predictions of
observables $\cal{R}$.
While phase space effects which are introduced by massive quarks can  often
be studied using hadronization models and Monte Carlo simulation techniques,
explicit quark mass corrections in higher than leading perturbative order are
available only for  
very few observables.

Secondly, any quark-mass dependence of $\cal{R}$ will add another term
$\mu^2 \frac{\partial m}{\partial \mu^2}\frac{\partial}{\partial m} {\cal R}$
to equation~\ref{eq-muindependence}, which leads to energy-dependent, running quark
masses, $m_q (Q^2)$, in a similar way as the running coupling $\as (Q^2)$ was
obtained.

In addition to these effects, $\as$ indirectly also depends on the quark masses,
through the dependence of the $\beta$ coefficients on the effective number of
quarks flavours, $N_f$, with $m_q \ll \mu$.
Constructing an effective theory for, say, ($N_f$-1) quark flavours which must be
consistent with the $N_f$ quark flavours theory at the heavy quark threshold
$\mu^{(N_f)} \sim {\cal O} (m_q)$, results in matching conditions for the $\as$
values of the ($N_f$-1)- and the $N_f$-quark flavours theories \cite{bernreuther}.

In leading and in next-to-leading order, the matching condition is
$\as^{(N_f-1)} = \as^{N_f}$.
In higher orders and the $\overline{MS}$ scheme, however, nontrivial matching
conditions apply \cite{bernreuther,larin,alphas-4loop}.
Formally these are, if the energy evolution of $\as$ is performed in $n^{th}$
order, of order ($n-1$).

The matching scale $\mu^{(N_f)}$ can be chosen in terms of the (running)
$\overline{MS}$ mass $ m_q ( \mu_q )$, or of
the constant, so-called pole mass $ M_q $.
For both cases, the relevant matching conditions are given in
\cite{alphas-4loop}. 
These expressions have a particluarly simple form for the 
choice\footnote{The results of reference~\cite{alphas-4loop} are also valid for
other relations between $\mu^{(N_f)}$ and $m_q$ or $M_q$, as e.g. $\mu^{(N_f)}
= 2 M_q$. 
For 3-loop matching, however, practical differences due to the freedom of this
choice are negligible.}
$\mu^{(N_f)} = m_q (m_q)$ or $\mu^{(N_f)} = M_q$.
In this report, the latter choice will be used to perform 3-loop matching
at the heavy quark pole masses, in which case the matching condition reads, with
$a = \as^{(N_f)} / \pi$ and $a' = \as^{(N_f-1)} / \pi$:

\begin{equation} \label{Mq-matching}
\frac{a'}{a} = 1 + C_2\ a^2 + C_3\ a^3 \ ,
\end{equation}
\noindent
where $C_2 = - 0.291667$ and $C_3 = -5.32389 + (N_f-1)\cdot 0.26247$
\cite{alphas-4loop}.

\begin{figure}[ht]
\begin{center}
\epsfxsize12.0cm\epsffile{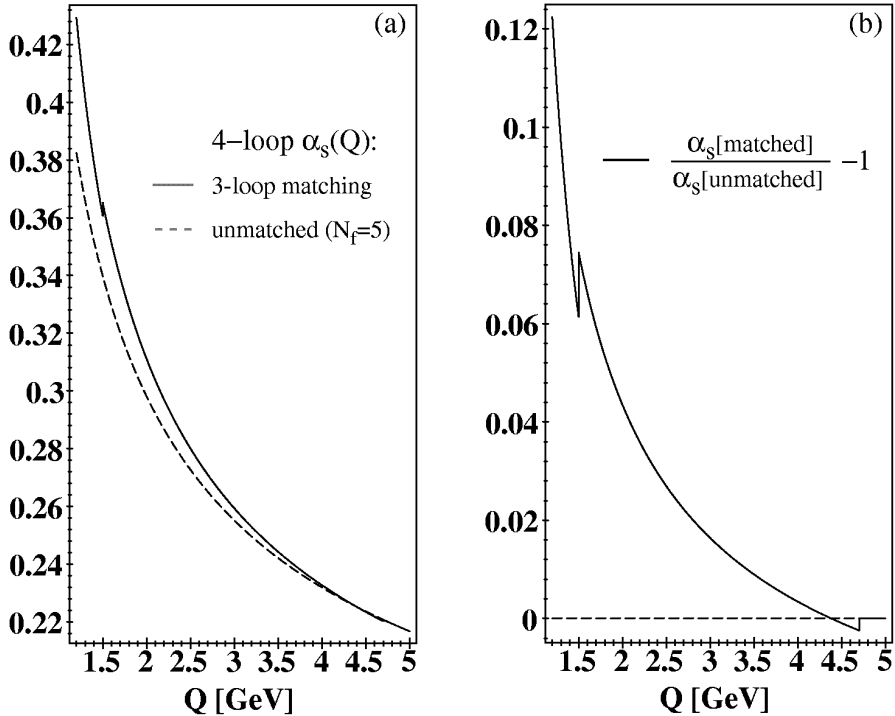} 
\end{center}
\caption{(a) 4-loop running of $\as (Q)$ with 3-loop quark threshold matching
according to equations~\protect\ref{eq-as4loop} and~\protect\ref{Mq-matching}, with
$\lamsb^{(N_f = 5)}$~=~220~MeV and charm- and bottom-quark thresholds at the
pole masses,
$\mu^{(N_f = 4)}_c \equiv M_c = 1.5$~GeV and $\mu^{(N_f = 5)}_b \equiv M_b =
4.7$~GeV (full line),
compared with the unmatched 4-loop result (dashed line).
(b) The fractional difference between the two curves in (a).
\label{fig:as-match}}
\end{figure}

The 4-loop prediction for the running $\as$, using equation~\ref{eq-as4loop} with
$\lamsb^{(N_f = 5)}$~=~220~MeV and 3-loop matching at the charm- and bottom-quark
pole masses,
$\mu^{(N_f = 4)}_c = M_c = 1.5$~GeV and $\mu^{(N_f = 5)}_b = M_b = 4.7$~GeV,
is illustrated in figure~\ref{fig:as-match}a (full line).
Small discontinuities at the quark thresholds can be seen, such that
$\as^{(N_f-1)} < \as^{(N_f)}$ by about 2~per mille at the bottom- and about
1~per cent at the charm-quark threshold.
The corresponding values of $\lamsb$ are $\lamsb^{(N_f=4)} = 305$~MeV and
$\lamsb^{(N_f=3)} = 346$~MeV.
Comparison with the 4-loop prediction, without applying
threshold matching and for $\lamsb^{(N_f = 5)}$~=~220~MeV and $N_f$~=~5 
throughout (dashed
line) demonstrates that, in addition to the discontinuities, 
the matched calculation
shows a steeper rise towards smaller energies because of the larger 
values of
$\lamsb^{(N_f=4)}$ and $\lamsb^{(N_f=3)}$.

The size of discontinuities and the changes of slopes are demonstrated
in figure~\ref{fig:as-match}b, where the fractional difference between the two
curves from figure~\ref{fig:as-match}a, i.e. between the matched and the unmatched
calculation, is presented.
Note that the step function of $\as$ is not an effect
which can be measured; the steps are artifacts of the truncated
perturbation theory and the requirement that predictions for observables at
energy scales around the matching point must be consistent and independent of the
two possible choices of (neighbouring) values of $N_f$.

\subsection{Perturbative predictions of physical quantities}

In practice, $\as$ is not an `observable' by itself.
Values of $\as (\mu^2)$ are
determined from measurements of observables $\cal{R}$ for which QCD
predictions exist. 
In perturbative QCD,
these are usually given by a power series in $\as (\mu^2 )$, like
\begin{eqnarray} \label{eq-rseries}
{\cal R}(Q^2) &=&   P_{l} \sum_{n} R_n \as^n \nonumber \\
              &=& P_l \left( R_0 + R_1 \as (\mu^2) + R_2 (Q^2 / \mu^2 ) \as^2
(\mu^2 ) + ...\right) \  ,
\end{eqnarray}
\noindent 
where $R_n$ are the $n_{th}$ order coefficients of the perturbation series and
$P_l R_0$ denotes the lowest-order value of $\cal R$.

For processes which involve gluons already in lowest order perturbation theory,
$P_l$ itself may include (powers of) $\as$. 
For instance, this happens in case of the hadronic decay
width of heavy Quarkonia, $\Gamma (\Upsilon \rightarrow ggg \rightarrow\
hadrons)$ for which $P_l \propto \as^3$. 
If no gluons are involved in lowest order, as e.g. in $\epem \rightarrow\
q\overline{q} \rightarrow \ hadrons$ or in deep inelastic scattering processes,
$P_l R_0$ is a constant and the usual choice of normalisation
is $P_l \equiv 1$.
$R_0$ is called the {\it lowest order} coefficient and $R_1$ is the
{\it leading order} (LO) coefficient.
Following this naming convention, $R_2$ is the {\it next-to-leading order}
(NLO) and $R_3$ is the {\it next-to-next-to-leading order}
(NNLO) coefficient.

QCD calculations in NLO perturbation theory are
available for many observables $\R$ in high energy particle reactions
like hadronic event shapes, jet production rates, scaling violations
of structure functions.
Calculations including the complete NNLO are
available for some totally inclusive quantities,
like the total hadronic cross section in $\epem \rightarrow\ hadrons$, moments
and sum rules of structure functions in deep inelastic scattering processes,
the hadronic decay widths of the $\z0$ boson, of the $\tau$ lepton
and of heavy quarkonia like the $\Upsilon$ and the $J/\Psi$.
The complicated nature of QCD, due to the process of gluon self-coupling and the
resulting large number of Feynman diagrams in higher orders of perturbation
theory, so far limited the number of QCD calculations in complete NNLO.

Another approach to calculating higher order corrections is
based on the
resummation of logarithms which arise from soft and collinear
singularities in gluon emission \cite{resummation}.   
Application of resummation techniques and appropriate matching with fixed-order calculations are further detailed e.g. in \cite{concise}.

\subsection{Renormalization scale dependence.}

The principal independence of a physical observable $\R$
from the choice of the renormalization scale $\mu$ was expressed in
equation~\ref{eq-muindependence}.
Replacing $\as$ by $\as (\mu^2)$, using
equation~\ref{eq-rge}, and inserting the perturbative expansion of $\R$ 
(equation~\ref{eq-rseries}) into equation~\ref{eq-muindependence} results, 
for processes with constant $P_l$, in
\begin{eqnarray} \label{eq-muindependence2}
0 = \mu^2 \frac{\partial R_0}{\partial \mu^2} 
    + \as (\mu^2) \mu^2 \frac{\partial R_1}{\partial \mu^2} \nonumber 
    &+& \as^2 (\mu^2) \left[ \mu^2 \frac{\partial R_2}{\partial \mu^2} -
    R_1 \beta_0 \right]  \nonumber \\
    &+& \as^3 (\mu^2) \left[ \mu^2 \frac{\partial R_3}{\partial \mu^2} -
    [R_1 \beta_1 + 2 R_2 \beta_0] \right] \nonumber \\
    &+& {\cal O} (\as^4) \ .
\end{eqnarray}
Solving this relation requires that the coefficients of $\as^n (\mu^2)$ vanish
for each order $n$.
With an appropriate choice of integration limits one thus obtains
\begin{eqnarray} \label{eq-R-mudependence}
R_0 &=& {\rm const.}\ , \nonumber \\
R_1 &=& {\rm const.}\ , \nonumber \\
R_2 \left(\frac{Q^2}{\mu^2}\right) &=& R_2 (1) - \beta_0 R_1 \ln
\frac{Q^2}{\mu^2}\ , \nonumber \\
R_3 \left(\frac{Q^2}{\mu^2}\right) &=& R_3 (1) - 
[ 2 R_2(1) \beta_0 + R_1 \beta_1
]  \ln \frac{Q^2}{\mu^2} + R_1\beta_0^2 \ln^2 \frac{Q^2}{\mu^2}
\end{eqnarray}
\noindent
as a solution of equation~\ref{eq-muindependence2}.

Invariance of the complete perturbation series
against the choice of the renormalization scale $\mu^2$ 
therefore implies that the
coefficients
$R_n$, except $R_0$ and $R_1$, explicitly depend on $\mu^2$.
In infinite order, the renormalization scale dependence of $\as$ and of the
coefficients $R_n$ cancel; in any finite (truncated) order, however, the
cancellation is not perfect, such that all realistic perturbative QCD
predictions include an explicit dependence on the choice of the
renormalization scale.

This dependence is most pronounced in leading order QCD
because $R_1$ does not explicitly
depend on $\mu$ and thus, there is no cancellation of
the (logarithmic) scale dependence of $\as (\mu^2)$ at all.
Only in next-to-leading and higher orders, the scale dependence of the
coefficients $R_n$, for $n \ge 2$, partly cancels that of $\as (\mu^2)$.
In general, the degree of cancellation improves with the inclusion of higher
orders in the perturbation series of $\R$.

Renormalization scale dependence 
is often used to test and specify uncertainties
of theoretical calculations and predictions 
of physical observables.
In most studies, the central value
of $\asmu$ is determined or taken for $\mu$ equalling the 
typical energy of the underlying scattering reaction, like e.g.
$\mu^2 = \ecm^2$ in $\epem$ annihilation, 
and changes of the result when varying this
definition of $\mu$ within \oq reasonable ranges"
are taken as systematic
uncertainties. 

There are several proposals of how to optimize or fix the
renormalisation scale, see e.g. \cite{stevenson,grunberg,blm,siggiscale}
Unfortunately, there is no common agreement of how to optimize
the choice of scales or how to
define the size of the corresponding uncertainties.
This unfortunate situation should be kept in mind when 
comparing and summarising results from
different analyses.

For more details and examples on renormalisation scale dependences
see e.g. \cite{concise}.

\subsection{Nonperturbative QCD methods}

At large distances or low momentum transfers, $\as$ becomes large and
application of perturbation theory becomes inappropriate.
Nonperturbative methods have therefore been developed to describe
strong interaction processes at low energy scales of typically
$Q^2 < 1$~GeV$^2$, such as the fragmentation of quarks and gluons
into hadrons (\oq hadronisation") and 
the absolute masses and mass splittings of mesons. 

{\em Hadronisation models} are used in Monte Carlo approaches
to describe the transition of quarks and gluons into hadrons.
They are based on QCD-inspired mechanisms like the
\oq string fragmentation" \cite{string,pythia} or 
\oq cluster fragmentation" \cite{herwig}.
Those models usually contain a number of free parameters 
which must be adjusted in order to reproduce experimental data
well.
They are indispensable tools not only for detailed QCD studies
of high energy collision reactions, but are also important to 
assess the resolution and acceptance of experimental setups.

{\em Power corrections} are an analytic approach
to approximate nonperturbative hadronisation effects by means of
perturbative methods, introducing a universal, non-perturbative parameter
$$\alpha_0 (\mu_I ) = \frac{1}{\mu_I } \int^{\mu_I}_0 {\rm d} k\ \as (k) $$
to parametrize the unknown behaviour of $\as (Q)$ below a certain
infrared matching scale $\mu_I $ \cite{powcor}.
Power corrections are regarded as an alternative approach to describe
hadronisation effects on event shape distributions, instead of using
phenomenological hadronisation models.

{\em Lattice Gauge Theory} is one of the most developed nonperturbative 
methods (see e.g. \cite{weisz,PRS}) and is used to calculate, for instance, 
hadron masses, mass splittings and QCD matrix elements.
In Lattice QCD, field operators are applied on a discrete, 4-dimensional
Euclidean space-time of hypercubes with side length $a$.

\section{Tests of Asymptotic freedom in $\epem$ annihilations}

While historical and early developments of QCD were mainly inspired 
by the
results from deep inelastic lepton-nucleon scattering experiments,
hadronic final state of $\epem$ annihilations developed into
a prime-tool for precision tests of QCD, due to the availability of
higher effective energies and the point-like and simple nature of the
primary particle reaction.
Studies of hadron jets and determinations of $\as$, in a wide range
of centre of mass energies and with increasing experimental
and theoretical precision, led to many signatures of
asymptotic freedom and, equivalently important, to the evidence for
colour charged gluons and the SU(3) gauge structure of QCD.

\subsection{Jet production rates}

The study of multijet event production rates in $\epem$ hadronic
final states gave direct evidence for the energy dependence
of $\as$ and for asymptotic freedom, before summaries of 
explicite measurements
of $\as$ were able to provide a convincing case
(c.f. figure~\ref{fig:as89}).
The basic idea was rather simple and straight-forward.
It was based on a definition of resolvable jets of hadrons
which could be applied both in perturbative QCD calculations 
and on measured hadronic final states:

Within the JADE jet algorithm \cite{jadejet1}, the scaled pair mass of
two resolvable jets $i$ and $j$, $y_{ij} = M_{ij}^2 / E_{vis}^2$,
is required to exceed a
threshold value $y_{cut}$, where $E_{vis}$ is the sum of the measured energies of
all particles of a hadronic final state - or, in theoretical
calculations, of all quarks and gluons.
In a recursive process,
the pair of particles or clusters of particles $n$ and $m$
with the smallest value of $y_{nm}$ is replaced  by  (or \oq recombined" into)
a single jet or cluster $k$ with four-momentum $p_k = p_n + p_m$, as long 
as $y_{nm} < \yc$.
The procedure is repeated until all pair masses $y_{ij}$
are larger than the jet resolution parameter $\yc$, and the remaining
clusters of particles are called jets.

\begin{figure}[t]
\begin{center}
\epsfxsize10.0cm\epsffile{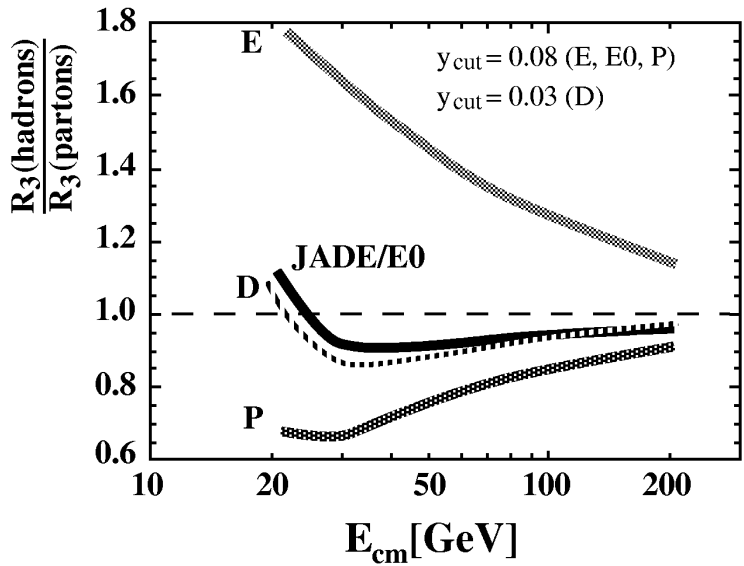} 
\end{center}
\caption{Energy dependence of hadronisation corrections to 3-jet
event production rates, defined using various jet reconstruction schemes.
\label{fig:r3}}
\end{figure}

Several jet recombination schemes and definitions of $M_{ij}$ exist
\cite{bkss};
the original JADE scheme with
$M_{ij}^2 = 2\cdot E_i\cdot E_j\cdot (1-\cos{\theta_{ij}})$, where $E_i$ and $E_j$
are the energies
of the particles and $\theta_{ij}$ is the angle between them,
and the \oq Durham" scheme \cite{durham}
with $M_{ij}^2 = 2\cdot {\rm min}( E_i^2, E_j^2)\cdot (1-\cos{\theta_{ij}})$, were
most widely used at LEP, due to their superior features like small sensitivity to
hadronisation and particle mass effects \cite{bkss}.

The virtues of the JADE jet definition were especially suited for an
early observation of the energy dependence of $\as$, without the need
to determine $\as$ and therefore avoiding the large systematic and 
- at that time - theoretical uncertainties:

\begin{itemize}
\item
The relative production rate $R_3$ of 3-jet hadronic final states
follows a particular simple theoretical expression:
$$ R_3 = \frac{\sigma (\epem \rightarrow 3-jets)}{\sigma 
(\epem \rightarrow hadrons)} = C_1\left( \yc \right) \asmu +
C_2(\yc ,\mu^2) \as^2(\mu^2) \ ,$$
in NLO perturbation theory.
In leading order, $R_3$ is thus directly proportional to $\as$,
and any energy dependence of $R_3$ observed in the data 
must be due to the energy dependence of $\as$ - if there are no other
energy dependent effects.
The coefficients $C_1$ and $C_2$ are energy independent.
They can be reliably calculated and
predicted by QCD, whereby the renormalisation scale dependence 
of $C_2$ is only a small disturbance.

\item
Model studies showed that hadronisation corrections to $R_3$ are
small and, in a suitable range of centre of mass energies, almost
constant, see figure~\ref{fig:r3} \cite{jadejet2}.

\item 
The JADE jet algorithm is particlarly easy  to apply to measured
hadronic final states, and corrections due to limited detector
resolution and acceptance are small and manageable.

\end{itemize}

\begin{figure}[ht]
\begin{center}
\epsfxsize12.0cm\epsffile{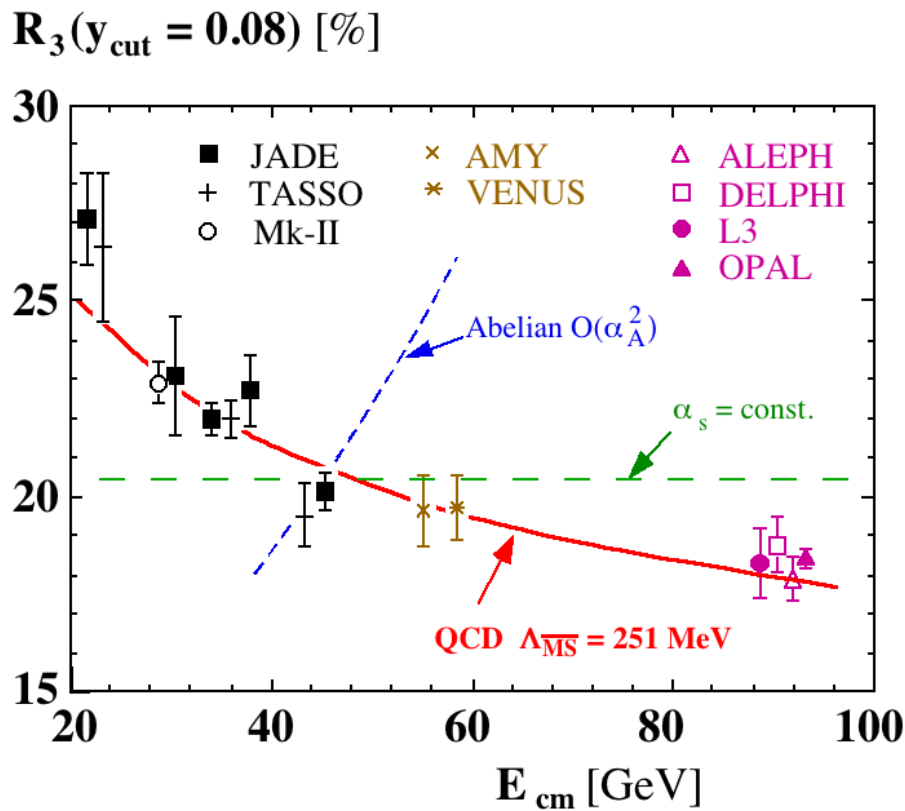} 
\end{center}
\caption{Energy dependence of 3-jet event production rates,
measured using the JADE jet finder at a scaled jet
energy resolution $\yc = 0.008$.
The errors are experimental.
The data are not corrected for hadronisation effects. 
They are compared 
to theoretical expectations of QCD, of an abelian vector gluon model,
and to the hypothesis of a constant coupling strength.
\label{fig:R3-08}}
\end{figure}

The first experimental study of the energy dependence of 3-jet
event production rates, at c.m. energies etween 22 and 46 GeV,
analysed for constant jet resolution
$\yc$ at the $\epem$ collider PETRA, 
gave first evidence for the energy dependence of $\as$
already in 1988 \cite{jadejet2}.
These data are shown in figure~\ref{fig:R3-08}, together with 
more results from eperiments at the PEP, TRISTAN \cite{sb-moriond-88}
and finally,
at the LEP collider \cite{sb-moriond-96}.
The measured 3-jet rates significantly decrease with 
increasing centre of mass energy, in excellent agreement
with the decrease predicted by QCD.
The hypthesis of an energy independent coupling, and especially
the prediction of an alternative, QED-like abelian vector gluon model,
where gluons carry no colour charge,
are in apparent contradiction with the data \cite{sb-moriond-96}.

\begin{figure}[ht]
\begin{center}
\epsfxsize12.0cm\epsffile{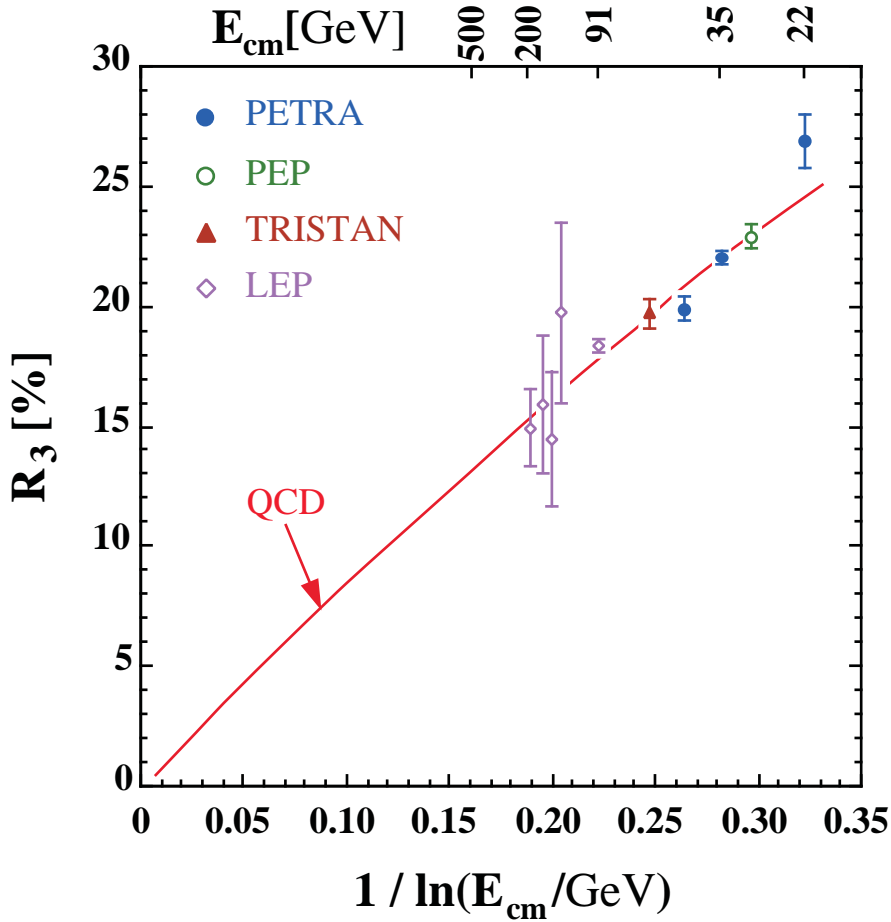} 
\end{center}
\caption{3-jet event production rates as shown in Fig.~\ref{fig:R3-08},
however as a function of $1 / \ln{\ecm}$, to demonstrate 
that $R_3 \propto \as \rightarrow 0$ at asymptotic (i.e. infinite) 
energies.
\label{fig:R3-lnQ}}
\end{figure}

In order to further demonstrate asymptotic freedom with these data,
they are - combined at suitable mean energies - plotted against
$1/\ln{\ecm}$, as shown in figure~\ref{fig:R3-lnQ}.
For infinite energies, $\ecm \rightarrow \infty$, $\as$ and thus $R_3$
are expected to vanish to zero, which is in very good agreement 
with the data.

\subsection{Evidence for the gluon self coupling}

The gluon self-coupling, as a direct consequence of gluons carrying
colour charge by themselves, is essential for the prediction of
asymptotic freedom. 
A rather direct method to detect effects of gluon-selfcoupling
was accomplished at the LEP collider, by analysing distributions
which are sensitive to the spin structure of hadronic 4-jet final states
\cite{sbzerwas}. 
For instance, the so-called Bengtson-Zerwas angle, $\chi_{BZ}$
\cite{bz},
measuring the angle between the planes defined by the two
highest and the two lowest energy jets, is rather sensitive to
the difference of a gluon-jet splitting into two gluons, which in QCD is the dominant source of 4-jet final states, and a gluon splitting
to a quark-antiquark pair, which is the dominant process in an
abelian vector theory where gluons carry no colour charge.

\begin{figure}[ht]
\begin{center}
\epsfxsize10.0cm\epsffile{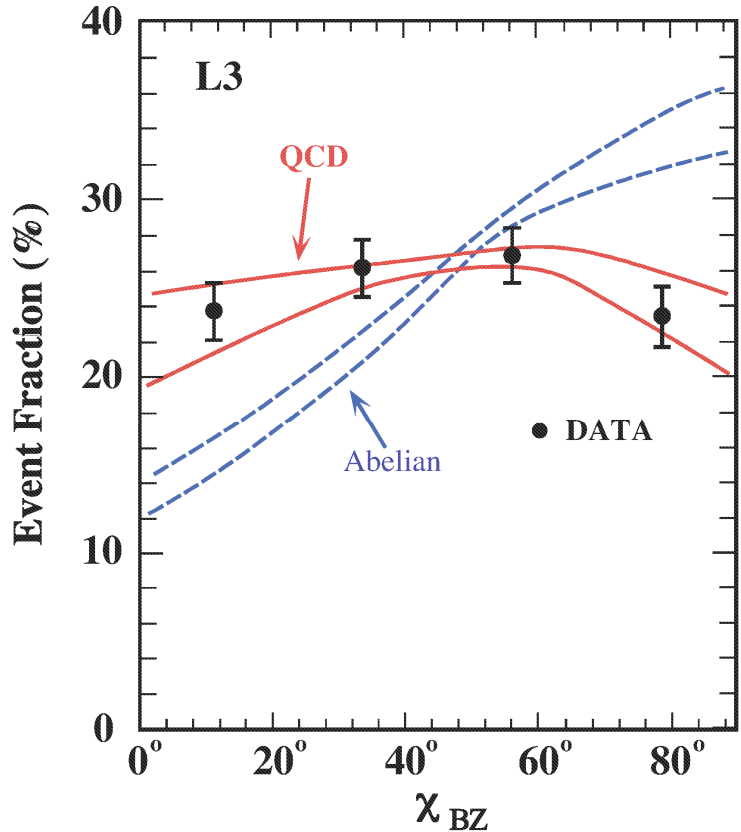} 
\end{center}
\caption{Distribution of the azimuthal angle between two planes
spanned by the two high- and the two low-energy jets of hadronic
4-jet events measured at LEP \cite{L3trip}, compared to the
predictions of QCD and of an abelian vector gluon model
where gluons carry no colour charge \cite{sbzerwas}.
\label{fig:L3trip}}
\end{figure}

The results of an early study which showed convincing evidence for
the gluon self coupling \cite{L3trip} after only one year
of data taking, 
is shown in figure~\ref{fig:L3trip}.
The data clearly favour the QCD prediction and rule out the abelian
vector gluon case.

\subsection{Determination of the QCD group constants}

As already mentioned in section~3, the QCD group constants
$C_A$, $C_F$ and $N_f$, assume values of 3, 4/3 and 5 in
a theory exhibiting SU(3) symmetry, with 5 quark flavours
in the vacuum polarisation loops.
For alternative, QED-like toy models of the strong
interaction with U(1) symmetry, these values would be 
$C_A=0$ and $C_F=0.5$.
Experimental determination of these two parameters can thus be regarded as
one of the most intimate tests of the predictions of QCD.

\begin{figure}[ht]
\begin{center}
\epsfxsize10.0cm\epsffile{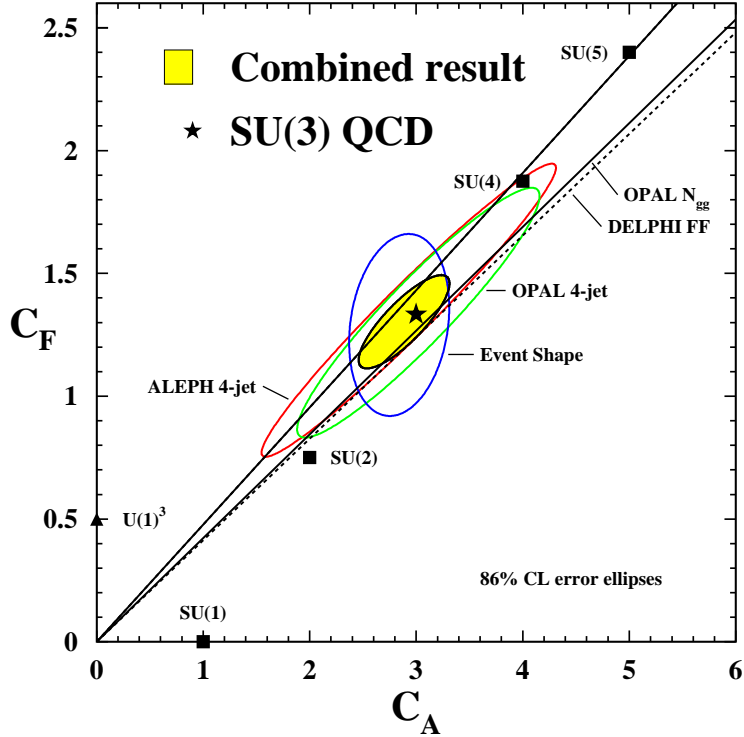} 
\end{center}
\caption{Measurements and combination of the QCD colour factors $C_A$ and $C_F$ \cite{kluth-tgv}.
\label{fig:cacfplot}}
\end{figure}

At LEP, data statistics and precision allowed to actually
determine experimental values for $C_A$, which basically is the 
number of colour charges, and $C_F$.
The current state-of-the art of such studies, involving  
analyses of several 4-jet angular correlations 
or fits to hadronic event shapes, is summarised \cite{kluth-tgv} 
in figure~\ref{fig:cacfplot}.
The data, with combined values of 
\begin{eqnarray}
C_A &=& 2.89 \pm 0.01\ {\rm (stat.)} \pm 0.21\ {\rm (syst.)} \\ \nonumber
C_F &=& 1.30 \pm 0.01\ {\rm (stat.)} \pm 0.09\ {\rm (syst.)}
\end{eqnarray}
are in excellent agreement with the gauge structure constants of 
QCD. 
They rule out the Abelian vector 
gluon model and theories exhibiting symmetries other than SU(3).

\subsection{The running $\as$ from 4-jet event production}

Since the beginning of the 1990's, advances in phenomenological 
calculations and predictions as well as in experimental techniques
allowed to perform precision determinations of $\as$, in a broad range
of energies and based on many different methods and observables.
While an overall summary of $\as$ measurements, as presented in 
section~6, gives a very distinct signature
for asymptotic freedom, the demonstration of the running of $\as$ from 
single experiments, minimising point-to-point systematic uncertainties,
adds extra confidence in the overall conclusion.

One of the most recent developments in determining $\as$ in $\epem$
annihilation is the precise extraction of $\as$ from differential 
4-jet distributions, i.e. the 4-jet production rate as a function
of the jet resolution $\yc$ \cite{a-4j,d-4j,o-4j}.
QCD predictions in NLO are available \cite{theo-4j}, which is
in $\oaaa$ perturbation theory.
4-jet final states  appear only at
$\oaa$ as their leading order, since they require at least two
radiation or splitting processes off the primary quark-antiquark
pair. 
Although \oq only" in NLO, the theoretical uncertainties for 4-jet
observables appear to be rather small, smaller than for typical 3-jet
event measures or event shapes. 
Since 4-jet observables are proportional to $\as^2$, they provide very 
sensitive measurements of $\as$.

As another new development, the data of the JADE experiment,
at the previous PETRA collider which was shut down in 1986,
are currently re-analysed, with the experimental and theoretical
experience and tools of today.
This adds important new information especially at lower energies, seen from the perspective of the LEP experiments. Moreover, the JADE experiment
can be regarded as the smaller brother or prototype of the OPAL 
detector at LEP, since both experiments were based on
similar detector techniques.
 
\begin{figure}[ht]
\begin{center}
\epsfxsize10.0cm\epsffile{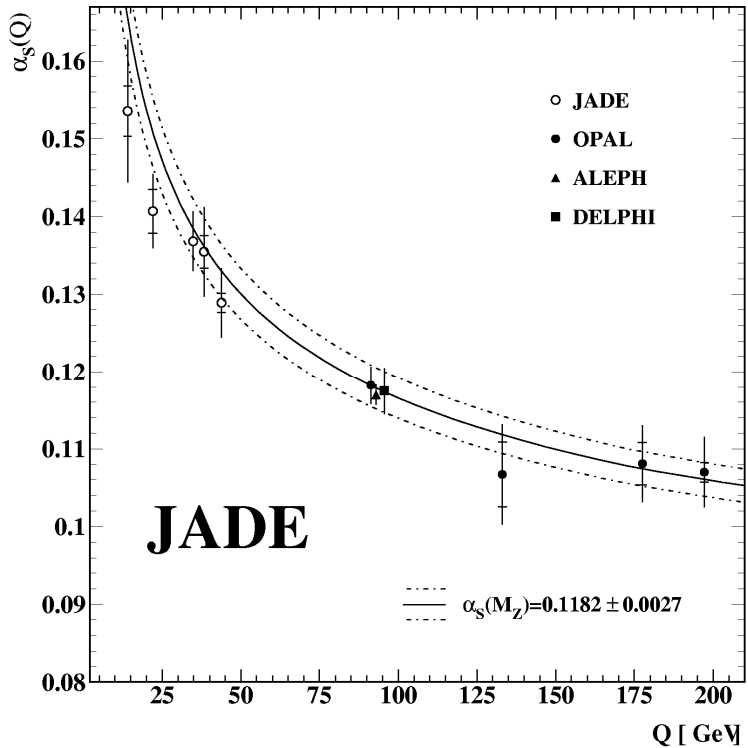} 
\end{center}
\caption{
The values of $\as$ from 4-jet event producion.
Errors are experimental (inner marks) and the total errors
\cite{j-4j,a-4j,d-4j,o-4j}.
The lines indicate the QCD prediction for the
running of $\as$ with $\amz = 0.1182 \pm 0.0027$ \cite{as-2004}.
\label{fig:asr4}}
\end{figure}

A LEP-like study of $\as$ from 4-jet production from JADE is available
\cite{j-4j}, and is summarised in figure~\ref{fig:asr4}, together
with the corresponding 
results from OPAL and the other LEP experiments.
In order to appreciate the high significance for the running
of $\as$, one should keep in mind that only the inner (experimental)
error bars need to be considered in a relative comparison of these data.

\section{QCD tests in deep inelastic lepton-nucleon scattering}

As outlined in section~2 and~3, the observation of approximate scaling 
of nuclear structure functions, and thereafter, with higher precision
and extended ranges of $x$ and $Q^2$, of (logarithmic) scaling 
violations, originally boosted the development of the quark-parton 
model and of QCD. 
The limited range of 
fixed-target lepton-nucleon scattering experiments in $x$ and $Q^2$,
however, prevented significant and unambiguous tests of QCD scaling
violations and the running of $\as$, see e.g. \cite{altarelli-89}.

This picture changed dramatically when the HERA
electron-proton and positron-proton collider started operation in 1991,
with lepton beam energies of 30 GeV and protons of 920 GeV.
HERA extended the
range in $Q^2$ by more than 2 orders of magnitude towards higher values,
and the range in $x$ by more than 3 orders of magnitude towards smaller
values.
With these parameters, precise tests of scaling violations
of structure functions, but also precise determinations of the
running $\as$
from jet production were achieved.
While these two topics will be reviewd in the following subsections,
a summary of significant $\as$ determinion in deep inelastic
scattering will be included in section~6, see also \cite{concise,as2002,
as-2004}.

\subsection{Basic introduction to structure functions}

Cross sections of physical processes in lepton-nucleon scattering and
in hadron-hadron collisions depend on the quark- and
gluon-densities in the nucleon.
Assuming factorisation between short-distance, hard scattering processes which
can be calculated using QCD perturbation theory, and low-energy or long-range
effects which are not accessible by perturbative methods,
such cross sections are 
parametrized by a set of structure functions $F_i$ ($i$= 1,2,3).
The transition between the long- and the short-range regimes is defined by an
arbitrary factorization scale $\mu_f$, which --- in general --- is independent
from the renormalization scale $\mu$, but has similar features as the latter:
the higher order coefficients of the perturbative QCD series for physical cross
sections depend on $\mu_f$ in such a way that the cross section to all orders
must be independent of $\mu_f$, i.e. 
$\mu_f \partial \sigma / \partial \mu_f = 0$.
To simplify application of theory to experimental measurements, the assumption
$\mu_f = \mu$ is usually made, with $\mu \equiv Q$ as the standard choice of
scales.

In the naive quark-parton model, i.e. neglecting gluons and QCD,
and for zero proton mass,
the differential cross sections for electromagnetic charged lepton
(electron or muon) - proton scattering 
off an unpolarized proton target is written
\begin{equation} \label{eq-dis}
\frac{{\rm d}^2 \sigma^{em}}{{\rm d}x {\rm d}Q^2} = \frac{4\pi \alpha^2}{Q^4}
\left[ [1 + (1 + y)^2 ] F_1^{em} + \frac{(1-y)}{x}(F_2^{em} - 2xF_1^{em})
\right]\ ,
\end{equation}
where, for fixed target reactions, $x = \frac{Q^2}{2M(E-E')}$
is the nucleon momentum fraction carried by the struck parton,
$y = 1 - E'/E$,
$Q^2$ is the negative quadratic momentum transfer in the scattering process, and
$M$, $E$ and $E'$ are the mass of the proton and the lepton energies before and
after the scattering, respectively, in the rest frame of the proton.
In the quark-parton model, these structure functions consist of combinations
of the quark- and antiquark densities $q(x)$ and $\overline{q}(x)$ for both
valence- (u,d) and sea-quarks (s,c).

\begin{figure}[ht]
\begin{center}
\epsfxsize14.0cm\epsffile{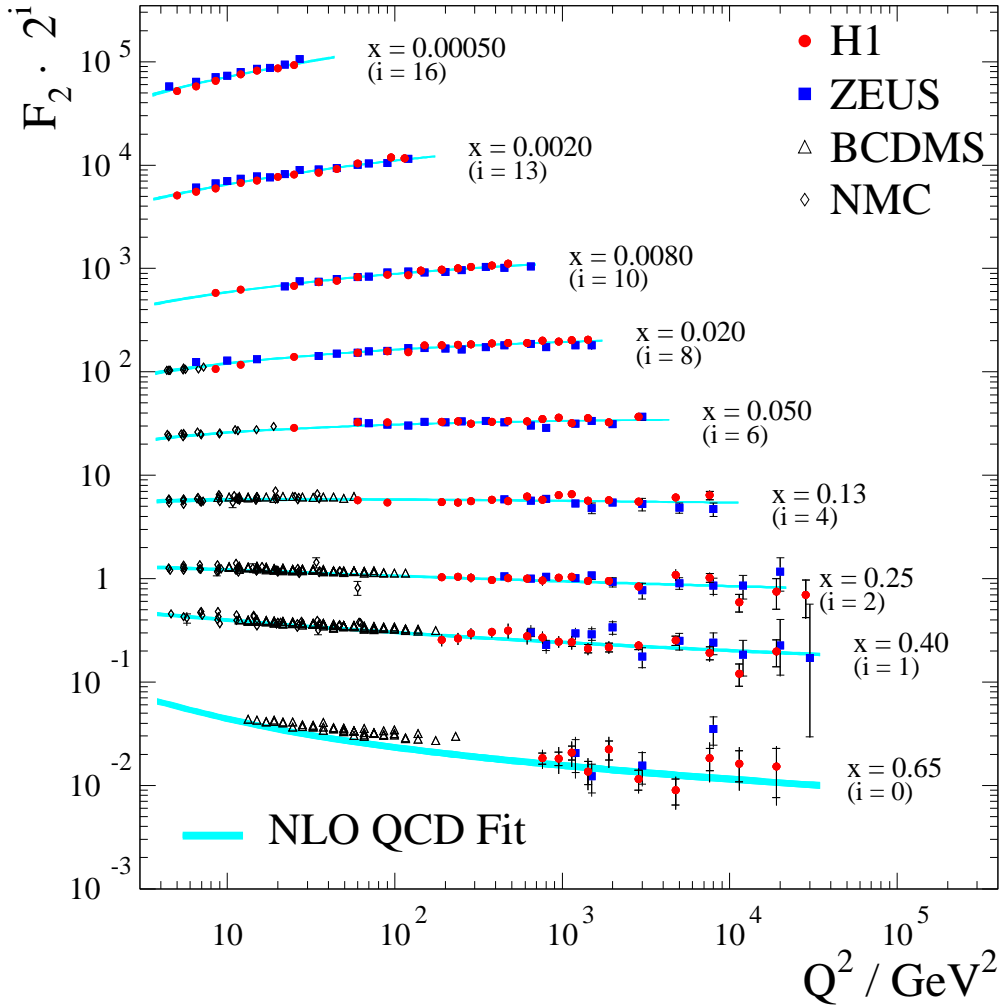} 
\end{center}
\caption{Summary of measurements of $F_2$ \cite{chekelian}.
For better visibility, the results for different values of $x$
were multiplied with the given factors of $2^i$.
\label{fig:f2}}
\end{figure}

In QCD, the gluon content of the proton as well as higher order diagrams describing
photon/gluon scattering, $\gamma g \rightarrow q \overline{q}$, 
QCD Compton processes $\gamma q \rightarrow g q$
and gluon radiation
off quarks must be taken into account.
Quark- and gluon-densities, the structure functions $F_i$ and physical
cross sections become energy ($Q^2$) dependent.
QCD thus predicts, departing from the naive quark-parton model,
scaling violations in physical cross sections, which are associated
with the radiation of gluons.
While perturbative QCD cannot predict the functional form of parton densities and
structure functions, their energy evolution is described by the
so-called DGLAP equations \cite{dglap,ap}.
 
The energy dependence of structure functions are known, since the 
early 1980's, in NLO QCD \cite{sf-nlo}.
Recently, parts of the NNLO predictions became available, see e.g.
\cite{sf-nnlo,vogt},
and should be completely known in NNLO fairly soon.
Apart from terms whose energy dependence is given by 
perturbative QCD, structure functions contain
so-called \oq higher twist" contributions (HT).
The leading higher twist terms are proportional to $1/Q^2$; they are
numerically important at low $Q^2 < \cal{O}$(few~GeV$^2)$ and at very large $x
\simeq 1$.

\subsection{Scaling violations of structure functions}

A recent summary of measurements of the 
proton structure function $F_2$
is given in figure~\ref{fig:f2} \cite{chekelian}.
The measurements cover an impressively large parameter space
in $x$ and $Q^2$. 
The degree of scaling violations predicted by QCD \cite{gross2},
namely a strong increase of $F^2$ with increasing $Q^2$ at small $x$,
and a decrease of $F_2$ at $x > 0.1$, is clearly 
reproduced by the data.
Over the whole kinematic range, the data are in very good 
agreement with the structure function evolution as predicted 
by QCD.

\begin{figure}[ht]
\begin{center}
\epsfxsize12.0cm\epsffile{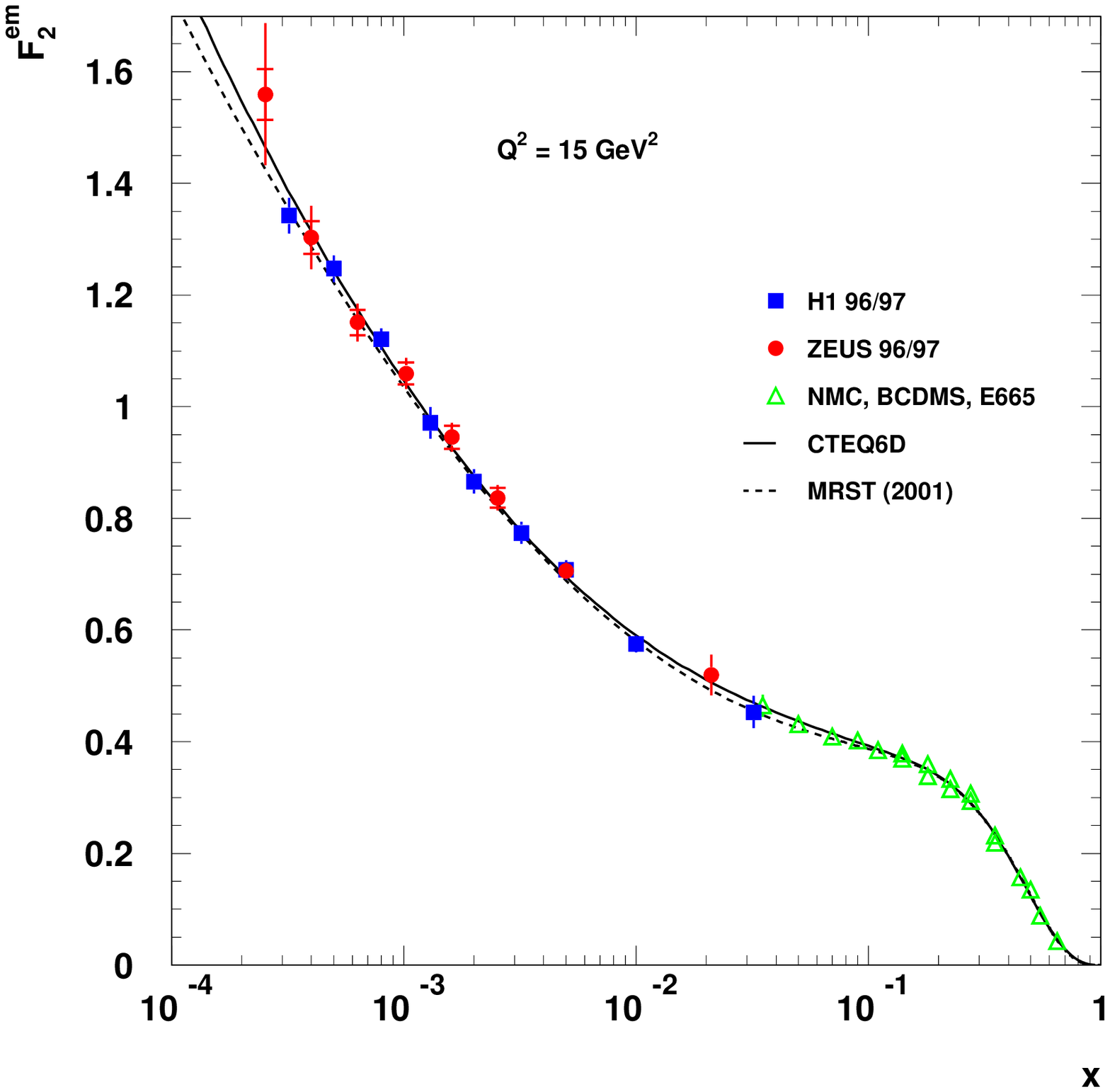} 
\end{center}
\caption{
Proton structure function $F_2(x,Q^2)$ \cite{chekelian}, as a function of $x$, 
demonstrating the strong rise at very small
values of $x$.
\label{fig:f2q15}}
\end{figure}

\begin{figure}[ht]
\begin{center}
\epsfxsize12.0cm\epsffile{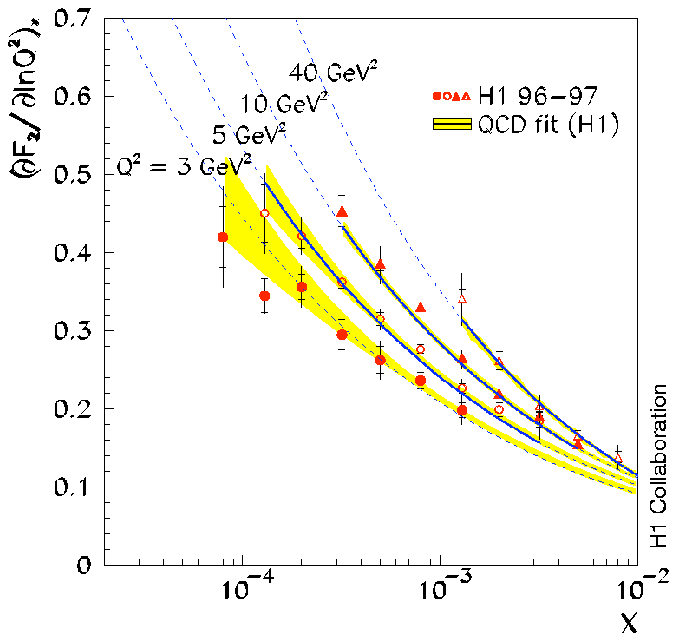} 
\end{center}
\caption{
Logarithmic slope of $F_2(x,Q^2)$ as a function of $x$ .
\label{fig:f2slopes}}
\end{figure}

One of the most important results obtained from HERA was, in fact, 
the discovery of the strong rise of $F_2$ at small $x$, as demonstrated
again in figure~\ref{fig:f2q15}.
It proves the QCD picture of an increasing part of the proton 
consisting of gluons, i.e. field energy, if the resolution
is increased (c.f. section~2.5).
Another way to demonstrate the compatibility of the observed
scaling violations with the predictions of QCD is to
analyse the lorgarithmic slopes, $\partial F_2(x,Q^2) / \partial ln Q^2$
as a function of $x$, as shown in figure~\ref{fig:f2slopes}.
Again, QCD provides an excellent description of the data.


\subsection{The running $\as$ from jet production in DIS}

The increase of momentum transfer $Q^2$ at HERA
also allowed to see and analyse the jet structure in deep inelasting
scattering processes. 
Here, similar as in $\epem$ annihilations, 
production of multijet final states is predicted in QCD to NLO,
and allows to determine $\as$ in a large range of energy scales.

Inclusive as well as differential jet production rates were studied in
the energy range of $Q^2\sim 10$ up to 10000~GeV$^2$, using similar
jet definitions and algorithms as in $\epem$ annihilation. 
In leading order $\as$,
2 + 1 jet events in deep inelastic $e p$ scattering arise from
photon-gluon fusion and from QCD Compton processes.
The term `2 + 1 jet' denotes events where two resolved jets can
be identified at large momentum transfer, 
in addition to the beam jet from the remnant of the 
incoming proton. 

\begin{figure}[ht]
\begin{center}
\epsfxsize14.0cm\epsffile{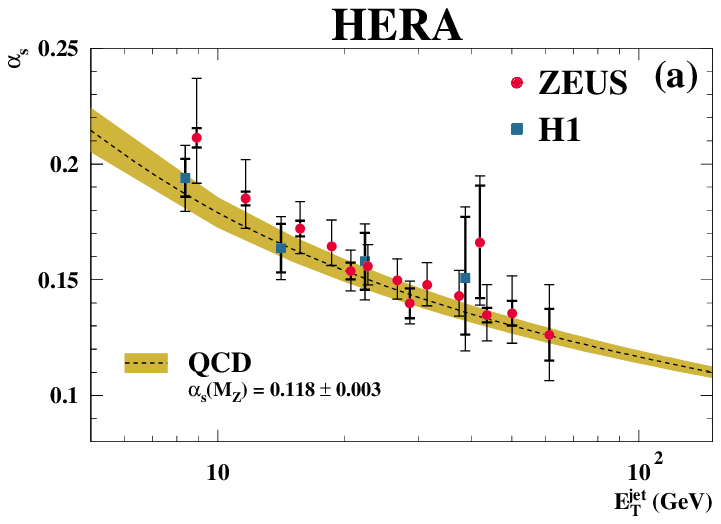} 
\end{center}
\caption{
Results of $\as$ as a function of $E_T^{jet}$ from HERA experiments H1
and ZEUS \cite{glasmann}.
\label{fig:as-hera-jets}}
\end{figure}

A recent summary of $\as$ determinations from the two HERA 
experiments H1 and ZEUS is given in figure~\ref{fig:as-hera-jets}
\cite{glasmann}.
Here, the transverse jet energy $E_T^{jet}$ was chosen as the relevant
energy scale. 
Both experiments have determined $\as$ at several different 
values of $E_T^{jet}$, and the summary of all these results
clearly demonstrates that $\as$ runs as predicted by QCD.

A combination of these results \cite{glasmann} will be included
in the overall summary of $\as$ determinations, which is presented in the
following section.

\section{Summary of $\as$ measurements}

\subsection{Previous developments and current status}

By the end of the 1980's, as concluded in section 2.11 above, 
summaries of measurements of $\as$ demonstrated very good
agreement with the prediction of asymptotic freedom (c.f. 
figure~\ref{fig:as89}), but both the precision and the energy range of
the results did not yet allow to claim firm evidence for the running
of $\as$.

This situation changed significantly with the turn-on of the LEP
$\epem$ collider in 1989, with advancements in theoretical
calculations and predictions, and with improved experimental
methods to determine $\as$.
At LEP, after the first successful year of data taking, precise
results emerged at high energies, i.e. at the $\z0$ mass peak, $\ecm = 
\mz = 91.2$~GeV, and also at very small energy scales of
$M_\tau = 1.78$~GeV, extending the energy range and improving the
overall precision by significant amounts.

This in turn quickly led to a very consistent picture: the experimental
results agreed with the prediction of asymptotic freedom in the
energy range from the $\tau$ to the $\z0$ mass, and were compatible
with a common value of $\amz = 0.117 \pm 0.004$ \cite{sb-catani}.
The theoretical uncertainties of this average were, and still are today, 
a matter of discussion.

The number of precise measurements of $\as$ continued to grow, 
especially from the HERA collider, from hadron collisions at the Tevatron, 
and from the study of heavy quarkonia masses and decays. 
Through LEP-II and the Tevatron, the energy scale of $\as$ determinations
was extended to 200~GeV and beyond.
The precision of many measurements and theoretical 
calculations was increased, and the number of observables 
and measurements
which were based on complete NNLO QCD predictions grew.
A comprehensive summary of the state-of-the-art 
of $\as$ determinations was given
e.g. in \cite{concise} and was continuously updated in
\cite{as2002,as-2004}, leading to a combined world average 
of $\amz = 0.1182 \pm 0.0027$ in 
2004\footnote{
For alternative summaries of $\as$, leading to almost identical
values of $\amz$ with a different treatment of the
overall uncertainties, see e.g. \cite{pdg}}.

In the following, these previous summaries will be updated, the most
recent and new results will be presented, the actual evidence for
asymptotic freedom will be demonstrated and a new world average
of $\amz$ will be derived. 

\subsection{Update: new results since 2004}

Since the last summary in 2004 \cite{as-2004}, a number of
new results and updates of previous studies were presented.
The most significant which will be included in the
overall summary and determination of the world average of $\amz$
were:
\begin{itemize}

\item
A re-evaluation \cite{davier} 
of the existing data on hadronic decays of the $\tau$-lepton
and a revision of the theoretical framework,
especially in terms of narrowing the range of theoretical uncertainties,
resulted in a significantly improved value of
$\as (M_\tau) = 0.345 \pm 0.010$, in full NNLO QCD.
The total error is thereby reduced by a factor of 3, w.r.t.  previous
estimates \cite{lepqcd04} - which is basically due to 
the inclusion of available terms in NNNLO and a rather 
restrictive treatment of systematic uncertainties.
Although the error on this result, with a wider treatment of
systematics, can well increase by a factor of~2 \cite{kluth-private}, 
the published value is retained for further analysis in this review.

\item 
New analyses and a new combination of results from jet production
in deep inelastic electron or positron - proton scattering at HERA
\cite{glasmann}, as shown in figure~16, 
provided an improved overall value of $\amz = 0.1186 \pm 0.0051$,
in NLO of perturbative QCD.

\item
A new study of hadron masses using predictions from lattice gauge theory,
including vacuum polarisation effects from all three light quark flavours
and improved third and higher order perturbative terms,
resulted in a new and improved value of
$\amz = 0.1170 \pm 0.0012$ \cite{lgt}.
Although the methods used in this study and the small size of the
claimed overall error are still under discussion \cite{weisz-private}, 
the published
value is retained here for further discussion.

\item
New studies of 4-jet final states in $\epem$ annihilation at LEP
\cite{a-4j,d-4j,o-4j,j-4j}, see also section~4.4,
and a combination of the respective 
$\as$ results give a new average of $\amz = 0.1176 \pm 0.0022$,
in $\oaaa$ which, 
for 4-jet production, corresponds
to NLO in perturbative QCD.

\end{itemize}

In the following overall summary of measurements of $\as$, these
four results will replace the respective values used in the 
previous summary of 2004 \cite{as-2004}.

\subsection{$\as$ summary}

The new overall summary of $\as$ is given
in table~\ref{tab:astab},
where the new and updated results discussed in the previous section
are underlined.
Most of the results given in table~\ref{tab:astab} are combined from 
several or many individual measurements of different experiments and
groups. 
For results obtained at fixed energy scales $Q$ (or in narrow ranges of $Q$),
the value of $\as (Q)$ is given, together with the extrapolation
to the \oq standard" energy scale, $Q = \mz$,
using equation~\ref{eq-as4loop} in 4-loop approximation
and 3-loop quark threshold matching at the heavy quark pole
masses $M_c = 1.5$~GeV and $M_b = 4.7$~GeV.
Results from data in ranges of energies are only given
for $Q = \mz$. 
Where available, the table also contains the contributions of
experimental and theoretical uncertainties to the total errors in
$\amz$.

Finally, in the last two columns of table~\ref{tab:astab}, the 
underlying theoretical calculation for each measurement 
and a reference to this result are given, where 
NLO stands for next-to-leading order, NNLO for next-next-to-leading-order
of perturbation theory, \oq resum" stands for resummend NLO calculations
which include NLO plus resummation of all leading und
next-to-leading logarithms to all orders (see \cite{resummation} 
and \cite{concise}), and \oq LGT" indicates lattice gauge theory.

\begin{figure}[ht]
\begin{center}
\epsfxsize13.0cm\epsffile{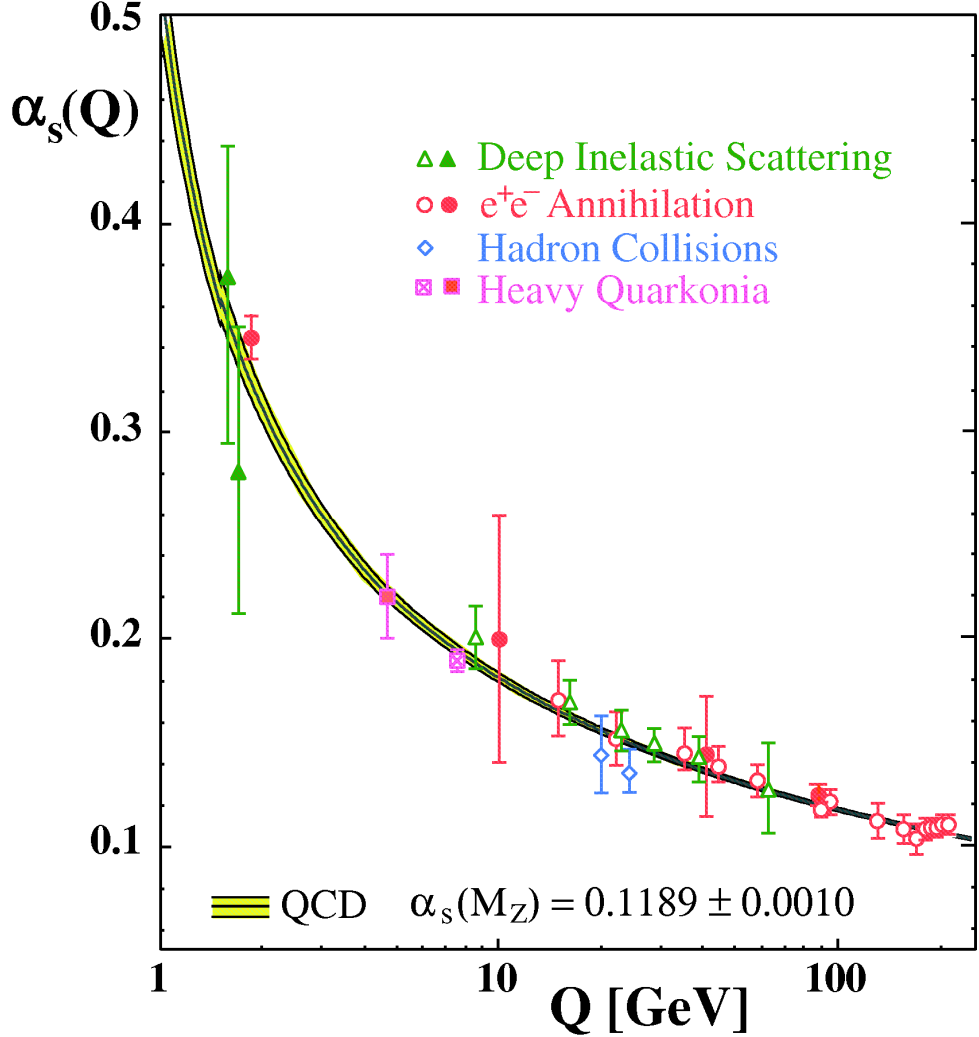} 
\end{center}
\caption{.
Summary of measurements of $\as (Q)$ as a function of the
respective energy scale $Q$, from table~\ref{tab:astab}.
Open symbols indicate (resummed) NLO, and
filled symbols NNLO QCD calculations used in the respective
analysis.
The curves are the QCD predictions for the combined world
average value of $\amz$, in 4-loop approximation and using 3-loop
threshold matching at the heavy quark pole masses
$M_c = 1.5$~GeV and $M_b = 4.7$~GeV.
\label{fig:asq-2006}}
\end{figure}

In figure~\ref{fig:asq-2006}, all results of $\as (Q)$ given in 
table~\ref{tab:astab} are graphically displayed, as a function of
the energy scale $Q$.
Those results obtained in ranges of $Q$ and given, in table~\ref{tab:astab},
as $\amz$ only, are not included in this figure - with one exception:
the results from jet production in deep inelastic scattering
are represented in table~\ref{tab:astab} by one line, averaging
over a range in $Q$ from 6 to 100~GeV, while in figure~\ref{fig:asq-2006}
combined results for fixed values of $Q$ as presented in \cite{glasmann}
are displayed.

The data are compared with the QCD prediction for the running $\as$,
calculated for $\amz = 0.1189 \pm 0.0010$ which - as will be discussed in the next subsection - is the new world average.
The QCD curves are calculated using the 4-loop perturbative prediction, equation~\ref{eq-as4loop}, and 3-loop quark threshold
matching, see section~3.6.
The data are in excellent agreement with the
QCD prediction, from the smallest to the largest energy scales 
probed by experimental data.

\begin{figure}[ht]
\begin{center}
\epsfxsize10.0cm\epsffile{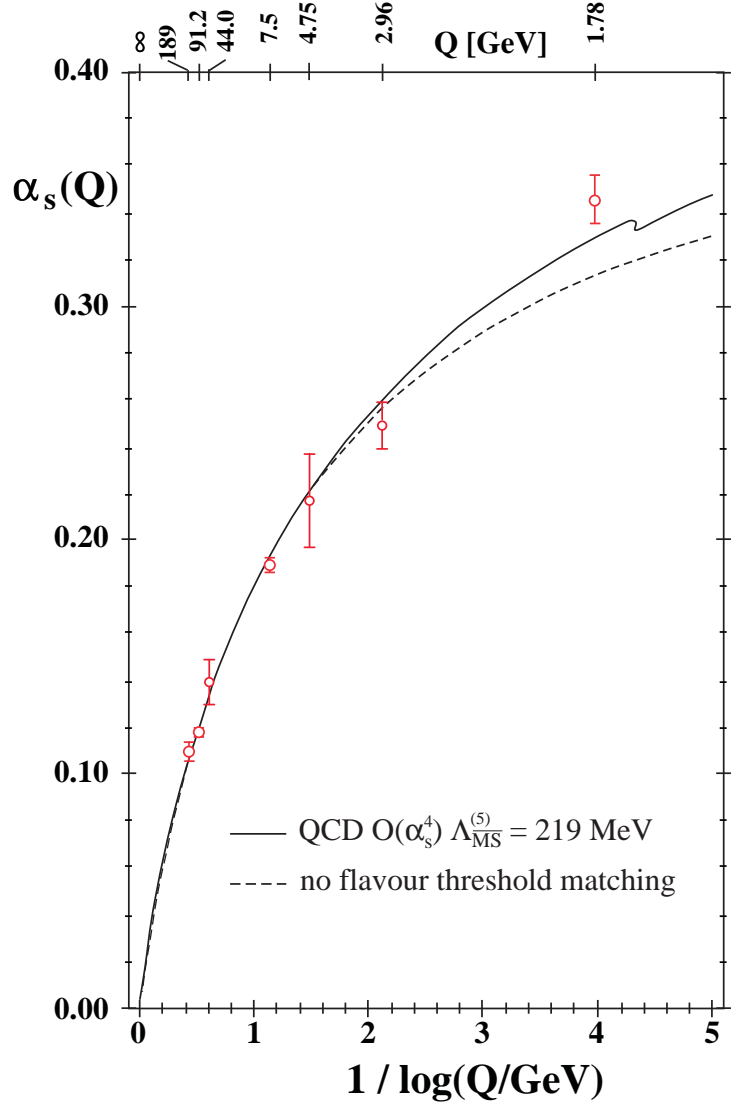} 
\end{center}
\caption{Some of the most significant data from figure~\ref{fig:asq-2006},
however plotted as a function of $1 / \log(Q / GeV)$, in order
to demonstrate the significance of the data showing evidence for
asymptotic freedom, i.e. the vanishing of $\as (Q)$ at asymptotically
large energy scales.
\label{fig:as06-logq}}
\end{figure}

In order to demonstrate the significance with which the data probe 
asymptotic freedom, some of the most precise $\as$ results are 
plotted\footnote{
Only a selection of results, especially of those at high energy
scales, is shown in this figure, to maintain visibility and clarity.
The measurements chosen for this demonstration, from left to
right, are:
$\as$ from $\epem$ event shapes and jets at LEP-II, for  ($Q$~=189~GeV),
from 4-jet events at LEP ($Q = \mz = 91.2$~GeV),
from the reanalysis of $\epem$ event shapes and jets at PETRA
($Q = 44$~GeV), 
from hadron masses and lattice theory at $Q = 7.5$~GeV,
from heavy quarkonia decays at $Q = 4.75$~GeV, and for
$\tau$-decays at $Q = 1.78$~GeV.} 
in figure~\ref{fig:as06-logq}, now as a function of the
inverse logarithm
of $Q$.
Also shown in this figure is the QCD prediction in 4-loop
approximation including 3-loop quark threshold matching,
for $\amz = 0.1189$ (full line), and,
for demonstration only, the 4-loop QCD
curve omitting quark threshold matching.

The data very significantly prove the particular QCD
prediction of asymptotic freedom.
Apart from precisely reproducing the characteristic 
QCD-shape with an inverse logarithmic slope, the data point at
the very lowest energies, i.e. the right-most point
in figure~\ref{fig:as06-logq},
indicates that the available precision allows
to conclude that quark threshold matching is necessary
for QCD to consistently describe the data.

In fact, data precision is now so advanced that a rather simple
QCD fit, e.g. in leading order QCD with no threshold matching,
with a fit probability of less than 1\%, fails to describe 
data\footnote{
Such a \oq simple" fit was previously used \cite{concise}
to \oq fit" the $\beta_0$ coefficient of the QCD beta-function,
c.f. equation~\ref{eq-betafunction}, or - alternatively - the number of
colour degrees of freedom, $C_A = N_c =3$.}.
Evidently, the probabilities for a hypothetical {\em constant} 
and energy $in$dependent $\as$\footnote{
In fact, there exists no theory which predicts a constant coupling.},
or an abelian vector gluon theory which predicts an $in$crease of
the coupling with increasing energy scale, c.f. figure~\ref{fig:R3-08},
have negligible probabilities to describe data.
The same is true for other functional forms, like
$\as \propto 1/Q$ or $\as \propto 1/Q^2$ - such functions may
be adjusted such that they can describe a few data points either at
low or at high values of $Q$, but altogether fail to describe data
in the full range of energy scales from 1.78 to 200 GeV.

Therefore, it is concluded that the data, with the current precision 
which has substantially increased over the past few years,
prove the specific QCD functional form of the running coupling $\as$,
and therefore of asymptotic freedom.

\subsection{A new world average of $\amz$}

Since all measurements of $\as$, as shown in figure~\ref{fig:asq-2006},
are consistent with the running coupling as predicted by QCD,
it is legitimate to evolve all values of $\as (Q)$ to a common
reference energy scale, like e.g. $Q = \mz$, using the 
QCD 4-loop beta-function.
This was already done in the past, see e.g. \cite{concise,as-2004}, 
and world average values of $\amz$ were determined.

A general difficulty in the averaging
procedure is the proper and optimal treatment
of errors and uncertainties.
For many of the results, the dominating error is the theoretical
uncertainty - which however is not uniquely defined, and is treated quite 
differently in most of the studies. 
While in a majority of analyses the theoretical uncertainty is 
defined through a variation of the renormalisation scale,
the range of this variation is not unique.
Theoretical errors, in some studies, are defined and probed 
in a rather restrictive and optimistic way, while in other
studies they are treated in a more generous manner.
Also, theoretical uncertainties are likely to be correlated
between different analyses, however to an unknown degree.
So the assigned errors cannot be treated like statistical and 
independent errors.

In previous studies \cite{concise,as2002,as-2004},
an error-weighted average and an \oq optimised correlation" was calculated
from the error covariance matrix, assuming an overall correlation
factor between the errors of the measurements included in the 
averaging process.
This factor was adjusted such that the overall $\chi^2$ equals unity
per degree of freedom \cite{schmelling}.
Usually the overall
$\chi^2$ was larger than unity per d.o.f., which indicated that either the
individual errors were overestimated, or they were correlated.
Adjusting an overall correlation factor should take care about 
the latter case.

The procedure, applied in a previous study
to all results which are based on NNLO
QCD and which have overall errors of $\le 0.008$ on $\amz$, then resulted
in $\amz = 0.1182 \pm 0.0027$ \cite{as-2004}.
Here, the averaging shall be done using the same method,
however the selection of results to be included in the final
average shall be altered:

First, the restriction to results which are complete to NNLO
will give a strong weight to studies at low energy scales, because
most of the results obtained at high energies are in NLO, possibly
including resummation, only.
Second, it may be time to include the results from lattice gauge 
calculations, which now have reached a maturity which may be
comparable to those in (NNLO) perturbation theory - and lattice gauge 
results are most likely $not$ correlated with classical perturbation
theory, thereby adding an independent input to the standard selection.
Finally, in order to reach a higher degree of independence between
the results included in the averaging procedure, data from different
processes should be chosen, avoiding a possible bias towards any
direction.
In this sense, the precise determinations of $\as$ from 4-jet
events in $\epem$ annihilation at $Q = M_z$, from event shapes and
jets at $Q = 189$~GeV and from jets in deep inelastic
scatterin are also chosen to be included.

Follwing this strategy, 10 results are henceforth 
selected to be included in the averaging procedure.
They are summarised in table~\ref{tab:assel}
and average to
$\amz = 0.1189 \pm 0.0007\ ,$
\noindent with an overall $\chi^2$ of 9.9 for 9 degrees of freedom.
Since the $\chi^2$ is larger than unity per d.o.f., no common 
correlation factor needs to be assumed; in fact, in order to reach exactly 
1 per d.o.f., the assigned errors of single measurements should be 
increased, a method which is frequently used e.g. in
\cite{pdg}.

\renewcommand{\arraystretch}{1.1}
\begin{table*}[ht]
{
\caption{
World summary of measurements of $\as$ (status of April 2006):
DIS = deep inelastic scattering; GLS-SR = Gross-Llewellyn-Smith sum rule;
Bj-SR = Bjorken sum rule;
(N)NLO = (next-to-)next-to-leading order perturbation theory;
LGT = lattice gauge theory;
resum. = resummed NLO. 
New or updated entries since the review of 2004 \cite{as-2004} are 
underlined.
\label{tab:astab}}
\begin{center}
\begin{tabular}{|l|c|c|c|c c|c|c|}
   \hline 
  & Q & & &  \multicolumn{2}{c|}
{$\Delta \amz $} & & \\ 
Process & [GeV] & $\alpha_s(Q)$ &
  $ \amz$ & exp. & theor. & Theory & refs.\\
\hline \hline 
DIS [pol. SF] & 0.7 - 8 & & $0.113\ ^{+\ 0.010}
  _{-\ 0.008}$ & $\pm 0.004$ & $^{+0.009}_{-0.006}$ & NLO &
\cite{bluemlein02}\\
DIS [Bj-SR] & 1.58
  & $0.375\ ^{+\ 0.062}_{-\ 0.081}$ & $0.121\ ^{+\ 0.005}_{-\ 0.009}$ & 
  -- & -- & NNLO & \cite{bjsr}\\
DIS [GLS-SR] & 1.73
  & $0.280\ ^{+\ 0.070}_{-\ 0.068}$ & $0.112\ ^{+\ 0.009}_{-\ 0.012}$ & 
  $^{+0.008}_{-0.010}$ & $0.005$ & NNLO & \cite{gls-recent}\\
\underline{$\tau$-decays} 
  & 1.78 & $0.345 \pm 0.010$ & $0.1215 \pm 0.0012$
  & 0.0004 &  0.0011 & NNLO & \cite{davier}\\
DIS [$\nu$; ${\rm x F_3}$]  & 2.8 - 11
  & 
   & $0.119\ ^{+\ 0.007}_{-\ 0.006}$   &
    $ 0.005 $ & $^{+0.005}_{-0.003}$ & NNLO & \cite{kataev2001}\\
DIS [e/$\mu$; ${\rm F_2}$]
     & 2 - 15 &      & $0.1166 \pm 0.0022$ & $ 0.0009$ &
     $ 0.0020$ & NNLO & \cite{yndurain2001,as2002}\\
\underline{DIS [e-p $\rightarrow$ jets]}
     & 6 - 100 &  & $0.1186 \pm 0.0051$ & $ 0.0011$ &
     $0.0050 $ & NLO & \cite{glasmann}\\
$\Upsilon$ decays
     & 4.75 & $0.217 \pm 0.021$ & $0.118 \pm 0.006
     $ & -- & -- & NNLO & \cite{Ydec-3rd}\\
\underline{${\rm Q\overline{Q}}$ states}
     & 7.5 & $0.1886 \pm 0.0032$ & $0.1170 \pm 0.0012 
     $ & 0.0000 & 0.0012
     & LGT & \cite{lgt}\\
$\epem$ [${\rm F^{\gamma}_2}$]
     & 1.4 - 28 &  & $0.1198\ ^{+\ 0.0044}_{-\ 0.0054}$ 
     & 0.0028 & $^{+\ 0.0034}_{-\ 0.0046}$ & NLO & \cite{lep-2gamma}\\
$\epem$ [$\sigma_{\rm had}$] 
     & 10.52 & $0.20\ \pm 0.06 $ & $0.130\ ^{+\ 0.021\ }_{-\ 0.029\ }$
     & $\ ^{+\ 0.021\ }_{-\ 0.029\ }$ & 0.002 & NNLO & \cite{cleo-rhad}\\
$\epem$ [jets \& shps]  & 14.0 & $0.170\ ^{+\ 0.021}_{-\ 0.017}$ &
   $0.120\ ^{+\ 0.010}_{-\ 0.008}$ &  0.002 & $^{+0.009}_{-0.008}$
   & resum & \cite{fernandez-2002}\\
$\epem$ [jets \& shps]  & 22.0 & $0.151\ ^{+\ 0.015}_{-\ 0.013}$ &
   $0.118\ ^{+\ 0.009}_{-\ 0.008}$ &  0.003 & $^{+0.009}_{-0.007}$
   & resum  & \cite{fernandez-2002}\\
$\epem$ [jets \& shps] & 35.0 & $ 0.145\ ^{+\ 0.012}_{-\ 0.007}$ &
   $0.123\ ^{+\ 0.008}_{-\ 0.006}$ &  0.002 & $^{+0.008}_{-0.005}$
   & resum  & \cite{fernandez-2002}\\
$\epem$ [$\sigma_{\rm had}$]  & 42.4 &
 $0.144 \pm 0.029$ &
   $0.126 \pm 0.022$ & $0.022
   $ & 0.002 & NNLO & \cite{haidt,concise}\\
$e^+e^-$ [jets \& shps] & 44.0 & $ 0.139\ ^{+\ 0.011}_{-\ 0.008}$ &
   $0.123\ ^{+\ 0.008}_{-\ 0.006}$ & 0.003 & $^{+0.007}_{-0.005}$
   & resum  & \cite{fernandez-2002}\\
$\epem$ [jets \& shps]  & 58.0 & $0.132\pm 0.008$ &
   $0.123 \pm 0.007$ & 0.003 & 0.007 & resum & \cite{as-tristan}\\
$\p\bar{\p} \rightarrow {\rm b\bar{b}X}$
    & 20.0 & $0.145\ ^{+\ 0.018\ }_{-\ 0.019\ }$ & $0.113 \pm 0.011$ 
    & $^{+\ 0.007}_{-\ 0.006}$ & $^{+\ 0.008}_{-\ 0.009}$ & NLO 
    & \cite{ua1-bb}\\
${\rm p\bar{p},\ pp \rightarrow \gamma X}$  & 24.3 & $0.135
 \ ^{+\ 0.012}_{-\ 0.008}$ &
  $0.110\ ^{+\ 0.008\ }_{-\ 0.005\ }$ & 0.004 &
  $^{+\ 0.007}_{-\ 0.003}$ & NLO & \cite{ppgam-recent}\\
${\sigma (\rm p\bar{p} \rightarrow\  jets)}$  & 40 - 250 &  &
  $0.118\pm 0.012$ & $^{+\ 0.008}_{-\ 0.010}$ & $^{+\ 0.009}_{-\ 0.008}$ & 
  NLO & \cite{cdf-jet}\\
$e^+e^-$ $\Gamma (\rm{Z \rightarrow had})$
    & 91.2 & $0.1226^{+\ 0.0058}_{-\ 0.0038}$ & 
    $0.1226^{+\ 0.0058}_{-\ 0.0038}$ &
   $\pm 0.0038$ & $^{+0.0043}_{-0.0005}$ & NNLO & \cite{lepew-0404}\\
\underline{{$e^+e^-$} 4-jet rate} 
&  91.2 & $0.1176 \pm 0.0022$ & $0.1176 \pm 0.0022$ & 
  0.0010 & 0.0020 & NLO & \cite{here}\\
$e^+e^-$ [jets \& shps] &
    91.2 & $0.121 \pm 0.006$ & $0.121 \pm 0.006$ & $ 0.001$ & $
0.006$ & resum & \cite{concise}\\
$\epem$ [jets \& shps]  & 133 & $0.113\pm 0.008$ &
   $0.120 \pm 0.007$ & 0.003 & 0.006 & resum & \cite{concise}\\
$\epem$ [jets \& shps]  & 161 & $0.109\pm 0.007$ &
   $0.118 \pm 0.008$ & 0.005 & 0.006 & resum & \cite{concise}\\
$\epem$ [jets \& shps]  & 172 & $0.104\pm 0.007$ &
   $0.114 \pm 0.008$ & 0.005 & 0.006 & resum & \cite{concise}\\
$\epem$ [jets \& shps]  & 183 & $0.109\pm 0.005$ &
   $0.121 \pm 0.006$ & 0.002 & 0.005 & resum & \cite{concise}\\
{$\bf e^+e^-$ [jets \& shps]} & 189 & $0.109\pm 0.004$ &
   $0.121 \pm 0.005$ & 0.001 & 0.005 & resum & \cite{concise}\\
$\epem$ [jets \& shps] & 195 & $0.109\pm 0.005$ &
   $ 0.122\pm 0.006$ & 0.001 & 0.006 & resum & \cite{as2002}\\
$\epem$ [jets \& shps] & 201 & $0.110\pm 0.005$ &
   $ 0.124\pm 0.006$ & 0.002 & 0.006 & resum & \cite{as2002}\\
$\epem$ [jets \& shps] & 206 & $0.110\pm 0.005$ &
   $ 0.124\pm 0.006$ & 0.001 & 0.006 & resum & \cite{as2002}\\
\hline
\end{tabular}
\end{center}
}
\end{table*}

\renewcommand{\arraystretch}{1.2}
\begin{table*}[ht]
{
\caption{
Measurements of $\amz$ included in the process to 
determine the world average, c.f. table~\ref{tab:astab}. 
The rightmost two columns give the exclusive mean value of
$\amz$ calculated {\em without} that particular measurement,
and the number of standard deviations between this
measurement and the respective exclusive mean,
treating errors as described in the text.
The inclusive average from {\em all} listed measurements
gives $\amz = 0.1189 \pm 0.0007$.
\label{tab:assel}}
\begin{center}
\begin{tabular}{|l|c|c||c|c|}
   \hline 
Process & Q [GeV] & $ \amz$ & excl. mean $\amz$ & std. dev.\\
\hline 
DIS [Bj-SR]          & 1.58     & $0.121\ ^{+\ 0.005}_{-\ 0.009}$  
 & $0.1189 \pm 0.0008$ & 0.3 \\
$\tau$-decays        & 1.78     & $0.1215 \pm 0.0012$              
 & $0.1176 \pm 0.0018$ & 1.8 \\
DIS [$\nu$; $x F_3$] & 2.8 - 11 & $0.119\ ^{+\ 0.007}_{-\ 0.006}$  
 & $0.1189 \pm 0.0008$ & 0.0 \\
DIS [e/$\mu$; $F_2$] & 2 - 15   & $0.1166 \pm 0.0022$              
 & $0.1192 \pm 0.0008$ & 1.1 \\
DIS [e-p $\rightarrow$ jets] & 6 - 100 & $0.1186 \pm 0.0051$       
 & $0.1190 \pm 0.0008$ & 0.1 \\
$\Upsilon$ decays    & 4.75     & $0.118 \pm 0.006$               
 & $0.1190 \pm 0.0008$ & 0.2 \\
${\rm Q\overline{Q}}$ states   & 7.5 & $0.1170 \pm 0.0012$       
 & $0.1200 \pm 0.0014$ & 1.6 \\
$\epem$ [$\Gamma (Z \rightarrow had)$ & 91.2 & $0.1226^{+\ 0.0058}_{-\ 0.0038}$ 
 & $0.1189 \pm 0.0008$ & 0.9 \\
$\epem$ 4-jet rate   &  91.2    & $0.1176 \pm 0.0022$             
 & $0.1191 \pm 0.0008$ & 0.6 \\
$\epem$ [jets \& shps] & 189    & $0.121 \pm 0.005$               
 & $0.1188 \pm 0.0008$ & 0.4 \\
\hline
\end{tabular}
\end{center}
}
\end{table*}

The fact that - in contrast to the case of previous reviews, see e.g.
\cite{concise,as-2004} - $\chi^2 / d.o.f.$ is not smaller than unity 
is mainly caused by two of the updates which were discussed above, namely
by the new assessment of $\tau$ decays and from heavy hadron masses
in lattice theory.
Both have the smallest overall errors assigned - $\pm 0.0012$ on 
$\amz$.
Treating these as gaussian and independent errors,
the two results are 2.7 standard deviations apart from each other.

The question whether one or both of these measurements has
underestimated its assigned overall uncertainty 
cannot finally be answered. 
The significance of each of the 10 selected measurements
can be judged from the last column of table~2, which shows 
the deviation of the respective average value of $\amz$ when $omitting$
this particular measurement in the averaging procedure.
The maximum deviation observed is $+0.0011 - 0.0013$,
a value which is compatible with a total error of $\pm 0.0010$.
The largest change in $\chi^2$ is observed when the result from $\tau$
decays is left out - a possible hint for an underestimation of its
assigned error.

Leaving out $two$ of the 10 selected measurements 
when averaging the results gives values of $\amz$ which vary between 
0.1173 and 0.1205; these two extremes again average to 0.1189
with a maximum deviation of 0.0016.

In view of these studies and variations, it is finally concluded that
$$\fbox{$\amz = 0.1189 \pm 0.0010$}$$
is the new world average\footnote{
A small increase of the error of 0.0007 from the covariance matrix
seems justified, due to the scatter of averages when leaving 
out one or two of the results, and due to $\chi^2$
being slightly larger than 1 per d.o.f.} of $\as$.
Here, the overall error decreased by almost a factor of three as compared
to the previous review \cite{as-2004}.
This small error, however, appears to be realistic since
all measurements agree well with this new average, as can be seen
in figure~\ref{fig:asq-2006}:
the error band is very narrow, but all the data are consistent with
this result.

We have therefore reached, after 30 years of QCD, the case that
not only asymptotic freedom is proven, beyond any doubts, by the
data, but also that $\amz$ is now known very precisely, to 
better than 1\% accuracy!

\section{Summary and Outlook}

The concept of asymptotic freedom, i.e. the QCD prediction
of an inverse logarithmic decrease of the coupling
strength $\as$ with the energy or the momentum transfer in
high energy scattering reactions, was shown to be significantly and 
reliably verified by a number of different measurements.
Historically, the first signatures for asymptotic freedom came from
the observation of approximate scaling of structure functions in
deep inelastic lepton-nucleon scattering experiments, and the
subsequent observation of small scaling violations, in a manner as 
was predicted by QCD.
Before specific measurements of $\as$ at different energy scales
and processes were precise enough to proove the running of
$\as$, measurements of the energy dependence of jet production rates in
$\epem$ annihilations provided strong evidence for 
the energy dependence of $\as$, as predicted by QCD and by 
the concept of asymptotic freedom.
The process of gluon self-coupling was experimentally established
by studies of angular correlations within 4-jet final states,
and the SU(3) gauge structure of QCD was confirmed by studies of
4-jet and event shape observables at the LEP $\epem$ collider.

Measurements of $\as$, performed by single experiments over ranges of
energy scales, 
demonstrate evidence for the running of $\as$, 
in perfect agreement with QCD.
Studies by single (or by few but similar) experiments
avoid systematic uncertainties which vary from 
(energy) point to point, and therefore have the potential to 
demonstrate asymptotic freedom, however, in most cases
only over a limited range of
energies. 
Examples of such studies, as $\as$ from jet production in deep inelastic
lepton-nucleon scattering, $\as$ from 4-jet event production 
in $\epem$ annihilation, are presented and discussed in this review.

The most significant experimental proof of asymptotic freedom today
is provided by the summary and combination
of all measurements of $\as$, over an energy range of 1.6 GeV to 
more than 200 GeV, from all available processes and experiments, 
involving perturbative and lattice QCD calculations.
The results are in excellent agreement with QCD and precisely
reproduce the inverse logarithmic dependence of $\as$ from the
energy or momentum transfer scale $Q$.
The data proove the neccessity to include higher loop diagrams
in the perturbative prediction of the energy dependence of $\as$,
and demonstrate the need for quark flavour threshold matching of
$\as$, again as predicted by QCD.

Since all measurements of $\as$ are in excellent agreement
with asymptotic freedom as predicted by QCD, 
the results are elvolved, according to the QCD 4-loop beta-function, 
to a common energy scale $Q \equiv \mz$.
A set of significant and precise measurements, well balanced
over all available processe, energies, experiments and 
theoretical methods, 
result in a new and improved world average value of
$$\amz = 0.1189 \pm 0.0010\ .$$
The overall uncertainty of this result is improved by almost a factor
of 3, compared to the previous average presented in 2004.

This improvement will have significant implications
on verifying the grand unification of forces,
and in particular, on possible signatures for Supersymmetry
see e.g. \cite{boer-sander,blair}.
The improvement
is achieved by the inclusion of several
new results and updates, like the one from $\tau$ decays,
from 4-jet final states in $\epem$ annihilations at LEP, and from 
lattice predictions of heavy hadron masses.

The total uncertainties of these new results are in the range
of 1~to~2~\% only, and therefore these measurements largely
dominate the determination of the world average.
Although some of the quoted uncertainties may be underestimated,
no significant disagreement between any
of the measurements nor with the world average value
of $\amz$ was found.

It is therefore concluded that the running coupling averages to the
value of $\amz$ given above, with an overall uncertainty now being
less than 1\%.

Future prospects for precise determinations of $\as$ are
upcoming QCD calculations, in NNLO perturbation theory, for more 
observables and processes, which will replace the currently used
calculations in NLO or in resummed NLO, see e.g. table~\ref{tab:astab}.
Such calculations are in preparation by several groups, 
see e.g. \cite{sf-nnlo,nnlo,vogt}, and the results are eagerly awaited
by experimentalists.
It is hoped that application of these new NNLO calculations will
provide an increased number of precise $\as$ determinations,
with overall errors of the order of 1\%. 
This will provide 
the means for precise compatibility checks and for further
improvements of the world average value of $\as$.

\eject

\noindent {\bf Apologies and Acknowledgements.} \\ \noindent
This review is based on an almost
countless number of studies and publications, comprising many 
hundreds of man-years of scientific and technical work on the
experimental setups, on running high- and low-energy 
accelerators and colliders, on developping the phenomenological
aspects of QCD,
not all of  which can be explicitly
referred to in this report.
I wish to thank all our colleagues who participated, in the past 30 years,
in the development of QCD, and in demonstrating its experimental
relevance. 
Apologies to those whose contributions did not receive the
appropriate attention in this report.
Special thanks go to Harald Fritzsch, Stefan Kluth, Peter Weisz 
and Peter Zerwas 
who were very helpful,
through many fruitful discussions, with their 
expertise and wisdom.

\end{document}